\documentclass[10pt,twocolumn]{article}

\usepackage[english]{babel}
\usepackage[T1]{fontenc}
\usepackage{lmodern}
\usepackage[letterpaper,margin=0.72in,columnsep=0.24in]{geometry}
\usepackage{amsmath,amssymb,amsfonts,bm}
\usepackage{mathtools}
\usepackage{graphicx}
\usepackage{xcolor}
\usepackage{booktabs}
\usepackage{tikz}
\usepackage[numbers,sort&compress]{natbib}
\usepackage{authblk}
\usepackage{titlesec}

\renewcommand{\thesection}{\Roman{section}}
\renewcommand{\thesubsection}{\Alph{subsection}}
\renewcommand{\thesubsubsection}{\arabic{subsubsection}}

\titleformat{\section}{\large\bfseries\raggedright}{\thesection}{0.75em}{}
\titleformat{\subsection}{\normalsize\bfseries}{\thesubsection}{0.75em}{}
\titleformat{\subsubsection}{\normalsize\itshape}{\thesubsubsection}{0.75em}{}

\usetikzlibrary{
  arrows.meta,
  calc,
  decorations.pathmorphing,
  patterns,
  positioning
}

\usepackage[colorlinks=true,
            linkcolor=blue,
            citecolor=blue,
            urlcolor=blue]{hyperref}

\newcommand{\R}{\mathbb{R}}

\newcommand{\drift}{\bm{v}}

\newcommand{\xvec}{\bm{x}}
\newcommand{\yvec}{\bm{y}}

\newcommand{\diff}{\mathrm{d}}

\newcommand{\br}{\boldsymbol{r}}
\newcommand{\bu}{\boldsymbol{u}}
\newcommand{\mk}{\mathcal{K}}

\newcommand{\bdry}{\partial\Omega}

\newcommand{\tsp}{\mathsf{\scriptstyle{T}}}

\begin{document}

\title{First-Hitting Location Laws as Boundary Observables of Drift--Diffusion Processes}

\author{Yen-Chi Lee\textsuperscript{*}}
\affil{Department of Mathematics, National Central University, Taoyuan, Taiwan}

\date{}

\twocolumn[
\begin{@twocolumnfalse}
\maketitle

\begin{center}
\footnotesize
\textcopyright\ 2026 American Physical Society. Published as
\textit{Phys. Rev. E} \textbf{114}, 024117 (2026);
\href{https://doi.org/10.1103/w1kn-fqfk}{doi:10.1103/w1kn-fqfk}.
\end{center}

\begin{abstract}
First-passage theory usually emphasizes when absorption occurs. Here, we instead
treat where absorption occurs---the first-hitting location (FHL)---as a primary
boundary observable of drift--diffusion processes. We formulate its law as the
exit measure induced by the diffusion generator and recover its density from the
normal derivative of an elliptic Green function. This yields exact
half-space kernels in arbitrary ambient dimension $d$ for constant drift, with the two- and
three-dimensional cases validated by Monte Carlo simulations. In the zero-drift
limit, the kernels reduce to scale-free Cauchy-type laws with algebraic tails;
drift introduces exponential screening and the characteristic length
$\ell_u=\sigma^2/\|\drift\|$, where $\sigma$ is the noise strength and $\drift$ is the constant drift vector, thereby localizing the boundary footprint. An
entropy-based effective width provides a finite diagnostic of this crossover.
To test the geometric reach of the planar theory, we also analyze
drift-free exterior hitting of a circle. Its exact Poisson kernel recovers the
planar Cauchy law as a local near-boundary limit, admits an exact Cauchy--Poisson
composition through an intermediate line, and yields a finite-time
supporting-line bound. These results organize geometry, drift, and pathwise
constraints within a unified description of first-hitting location laws.
\end{abstract}
\end{@twocolumnfalse}
]
\begin{NoHyper}
\begingroup
\renewcommand{\thefootnote}{*}
\footnotetext{\raggedright Corresponding author: yclee@math.ncu.edu.tw.}
\endgroup
\end{NoHyper}

\section{Introduction}
First-passage phenomena are central to nonequilibrium statistical physics,
stochastic transport, and reaction--diffusion theory
\cite{redner2001guide,berg2025random,bressloff2014stochastic,risken1989,bray2013persistence}.
They are usually characterized by temporal observables: the time to reach a
target, the survival probability, or the probability of absorption. For an
extended absorbing boundary, however, the location of the first contact is a
distinct physical observable. This first-hitting location (FHL) records how
transport dynamics and boundary geometry jointly organize absorption along an
interface, membrane, or detector \cite{holcman2014narrow}.

A concrete physical setting for this boundary-resolved observable
is engineered molecular communication, in which chemical or particulate
carriers transport information to a physical receiver
\cite{Akyildiz2008,farsad2016}. In diffusion-based architectures, the receiver
is often idealized as an absorbing surface, so that first contact constitutes
the reception event. Pandey, Mallik, and Lall made the spatial coordinate of
this event explicit by introducing the first-arrival-position channel
\cite{pandey2019}. For an extended, segmented, or geometrically structured
receiver, the resulting physical question is therefore not only whether or
when a carrier is captured, but where on the receiver the capture occurs and
how drift and boundary geometry shape that spatial pattern. The present work
addresses this boundary-resolved reception problem at the level of the
underlying stochastic transport law, without committing to a particular
modulation or detector architecture.

The FHL distribution is a classical probabilistic object. Without drift, it is
harmonic measure, represented by a Poisson kernel; with drift, it is the elliptic
measure associated with the diffusion generator and Dirichlet boundary
conditions \cite{doob2001classical,morters2010brownian,sapoval1993}.
Nevertheless, transport studies often emphasize stopping times or integrated
boundary fluxes, leaving the normalized spatial exit law implicit
\cite{risken1989,redner2001guide,grebenkov2007spectral}. Our aim is not to
redefine this classical measure, but to organize it as a boundary observable in
which the roles of geometry and drift are explicit.

We therefore combine two equivalent descriptions. The generator and its
Dirichlet Green function identify the FHL law as an induced boundary measure,
while the normal derivative of that Green function gives the familiar
boundary-flux density. For the diffusion models considered here,
the elliptic problem is the ensemble-level counterpart of the underlying
trajectory dynamics; their agreement for the present Dirichlet benchmark is
therefore a consistency relation, not an identification of their analytical
roles. The elliptic representation supports closed-form
calculation, whereas the stochastic formulation retains the stopping-event and
strong-Markov structure needed for geometric composition. This common language
allows the two viewpoints to be used complementarily rather than as competing
routes.

Our main drift-dependent benchmarks are planar absorbing boundaries. For
constant drift, we derive exact kernels in arbitrary ambient dimension $d$ and
specialize them to two and three spatial dimensions for numerical validation.
The drift-free laws have scale-free Cauchy-type tails. Drift reorganizes these
fluctuations through exponential screening and introduces the length
$\ell_u=\sigma^2/\|\drift\|=1/u$, with $u:=\|\drift\|/\sigma^2$, which separates diffusion-dominated and
drift-localized regimes. An entropy-based effective width is retained as a
compact diagnostic of this crossover.

To probe what survives beyond the half-space, we complement the
planar analysis with an exactly solvable drift-free circular receiver. This
benchmark distinguishes the local planar limit from global curvature, and it
exposes exact entrance-law and supporting-line roles for the point-to-line
kernel.

\paragraph{Main contributions.}
Because the Green-function and exit-measure representations are closely related to classical potential-theory concepts, it is important to clarify what is synthesized here for the stochastic transport and statistical physics communities. The primary advance of this work lies in systematically treating the FHL itself as the primary physical observable and organizing its stochastic and elliptic descriptions within a common framework. Specifically, we highlight the following contributions:
\begin{itemize}
\item \textbf{Stochastic--elliptic correspondence:} We provide a unified derivation that cleanly separates the spatial boundary measure from temporal stopping-time problems and makes explicit its equivalence to the corresponding elliptic boundary-flux formulation. The exponential tilting conversion, mathematically equivalent to a constant-drift Girsanov transformation \cite{oksendal2003}, connects the drifted generator to a modified Helmholtz problem, while the stochastic formulation additionally exposes pathwise and strong-Markov composition principles.
\item \textbf{General $d$-dimensional closed-form kernels:} We derive explicit, analytically closed-form half-space Poisson kernels for constant drift in arbitrary ambient dimension $d$. This result extends the explicit low-dimensional evaluations discussed in Ref.~\cite{polyanin2001handbook} to arbitrary ambient dimension.
\item \textbf{Thermodynamic regularization and numerical validation:} We identify a drift-induced crossover between scale-free diffusion and exponentially screened transport regimes. We explicitly demonstrate how directed transport acts as a localized probe, physically compressing scale-free Cauchy heavy tails into an exponentially screened spatial footprint. These analytical phenomena are explicitly validated by particle-based Monte Carlo simulations of the underlying Langevin dynamics.
\item \textbf{Curved-boundary benchmark and local planar limit:} For drift-free transport toward a circular absorbing receiver, we use the classical exterior Poisson kernel to show that its near-boundary scaling limit is the planar Cauchy law and to derive an exact Cauchy--Poisson composition through an intermediate line. A supporting-line construction further yields a finite-time upper bound for absorption by the convex receiver.
\end{itemize}
A concise summary of these methodological shifts compared to the standard parabolic PDE route is provided in Table~\ref{table:compare}. A compact dictionary of the corresponding stochastic and transport terminology is given in Table~\ref{table:dictionary}.

\paragraph{Scope and limitations.}
The main drift-dependent analysis focuses on planar absorbing geometries as canonical benchmarks for boundary-induced first-hitting location statistics.
To test which parts of this structure persist beyond a half-space, we also examine the drift-free exterior-circle problem as an exactly solvable curved-boundary benchmark.
This example isolates local curvature and global-shape effects without claiming a closed-form solution for arbitrary curved domains.
General curved boundaries with drift, time-dependent drift fields, and fully information-theoretic performance measures such as channel capacity remain beyond the scope of this study.

\paragraph{Paper organization.}
The remainder of the paper is organized as follows.
Section~\ref{sec:model} introduces the stochastic transport model and defines the first-hitting location observables considered in this work.
Section~\ref{sec:generator} develops a generator--Green-function framework that provides the analytical foundation for characterizing boundary exit laws via elliptic boundary-value problems.
In Sec.~\ref{sec:case}, explicit boundary kernels are derived for planar absorbing geometries in two and three dimensions, serving as canonical benchmarks for boundary-induced exit statistics.
Section~\ref{sec:information} examines the asymptotic structure of these exit laws and geometric diagnostics of their spatial spread.
Section~\ref{sec:circle} then develops an exactly solvable curved-boundary benchmark and its supporting-line consequences, and Sec.~\ref{sec:conc} concludes the paper.

\section{Model}
\label{sec:model}
The formulation adopted here is closely related to generator-based descriptions
of stochastic transport with absorbing boundaries \cite{doob2001classical,grebenkov2007spectral}.
Related approaches have also appeared in the molecular communication literature
\cite{farsad2016,pandey2019}.
Here, we isolate the transport layer underlying such receiver
architectures. Rather than fixing a particular modulation, readout, or decision
protocol, we ask how geometry and drift determine the spatial law of capture at
the boundary. This geometry-forward formulation preserves the concrete
interpretation of the FHL as a receiver observable while separating universal
transport effects from architecture-specific signal processing. The ideal
absorbing boundary is used as a minimal analytical benchmark in which this
transport law and its two representations can be established exactly.

\paragraph*{Notation and dimensional conventions.}
Throughout the paper, $d$ denotes the ambient spatial dimension.
We work in an ambient space $\mathbb{R}^d$ with
$d\ge 2$.
Particle trajectories evolve in a domain $\Omega\subset\mathbb{R}^d$ with
absorbing boundary $\partial\Omega$.
The boundary is therefore a $(d-1)$-dimensional manifold, and boundary-parallel
coordinates are denoted by $\boldsymbol{r}\in\mathbb{R}^{d-1}$.
For a planar half-space, $\lambda>0$ denotes the initial normal distance from the absorbing boundary.

The diffusion coefficient is denoted by $D>0$, with the standard parametrization
$D=\sigma^2/2$, following the conventional usage in statistical physics whereby
$D$ characterizes the macroscopic spreading rate of the process and appears as
the prefactor of the Laplacian in the diffusion equation.
In the equivalent stochastic differential equation formulation, the microscopic
noise amplitude enters via $\mathrm{d}X_t=\sigma\,\mathrm{d}B_t$, and the two
descriptions are related precisely through $D=\sigma^2/2$.
Accordingly, throughout this work we reserve the term \emph{diffusion coefficient}
for $D$, while referring to $\sigma$ as the noise strength.
For later use, it is convenient to introduce the inverse-length drift vector
\begin{equation}
\boldsymbol{u}:=\drift/\sigma^2,
\quad
\boldsymbol{u}=(\boldsymbol{u}_{\parallel},u_d),
\end{equation}
which naturally appears in the boundary kernel.
With this convention, the cases $d=2$ and $d=3$ correspond respectively to an
absorbing line and an absorbing plane, while higher-dimensional extensions fit
within the same unified framework.

\subsection{Stochastic transport model}
Let $\{\mathbf{X}_t\}_{t\ge 0}$ be an It\^{o} diffusion \cite{oksendal2003} in $\Omega$ with
absorbing boundary $\partial\Omega$.
The process starts from $\mathbf{X}_0=\xvec_0\in\Omega$ and satisfies the evolution equation:
\begin{equation}
\diff \mathbf{X}_t
=
\drift\,\diff t
+
\sigma\,\diff \mathbf{B}_t,
\quad t\ge 0,
\label{eq:IP}
\end{equation}
where $\mathbf{B}_t$ is a $d$-dimensional standard Brownian motion.
In the representative geometries studied in
Sec.~\ref{sec:case}, we consider constant drift $\drift$ and isotropic diffusion
with noise strength $\sigma$.

At the level of densities, the transition probability density
$p(\xvec,t\mid \xvec_0)$ of the killed process in $\Omega$
satisfies the advection--diffusion equation
with absorbing boundary condition $p=0$ on $\partial\Omega$ \cite{risken1989}:
\begin{equation}
\partial_t p
=
-\nabla\!\cdot(\drift\,p)
+
\frac{\sigma^2}{2}\,\Delta p,
\quad \mathbf{x}\in\Omega.
\label{ad-eq}
\end{equation}

\subsection{First-hitting observables}
Define the exit time from the domain $\Omega$ by
\begin{equation}
\tau_{\Omega}
:=
\inf\{t>0:\mathbf{X}_t\in\partial\Omega\},
\label{eq:hitting-time}
\end{equation}
and the corresponding boundary exit location by
\begin{equation}
\Xi := \mathbf{X}_{\tau_{\Omega}}\in\partial\Omega.
\end{equation}
The primary object of interest in this work is the \emph{first-hitting location}, viewed as a boundary-induced observable.
While boundary-hitting phenomena are often characterized through temporal
quantities such as survival probabilities or first-passage times, we instead
treat the normalized distribution of the boundary exit location as the central
observable \cite{redner2001guide,bray2013persistence,majumdar2007}.
From this perspective, geometry and drift act directly on the induced boundary
measure, rather than entering only through auxiliary stopping-time statistics.

Let $\omega^{\xvec}$ denote the exit measure on $\partial\Omega$ associated with
a trajectory starting from $\xvec$.
When $\partial\Omega$ is sufficiently regular, $\omega^{\xvec}$ is absolutely
continuous with respect to the surface measure $\diff S_{\yvec}$ on
$\partial\Omega$ \cite{doob2001classical,pinsky1995positive,makarov1985distortion}, and we write
\begin{equation}
\omega^{\xvec}(\diff \yvec)
=
\mk(\xvec,\yvec)\,\diff S_{\yvec},
\quad \yvec\in\partial\Omega,
\label{eq:fhl-def}
\end{equation}
where $\mk(\xvec,\yvec)$ is the associated boundary kernel, denoted throughout
by $\mathcal{K}(\xvec,\yvec)$ to avoid confusion with the modified Bessel function $K_\nu(\cdot)$.

Equivalently, for any test function $g$ defined on $\partial\Omega$, we have
\begin{equation}
\mathbb{E}_{\xvec}\!\left[g(\mathbf{X}_{\tau_{\Omega}})\right]
=
\int_{\partial\Omega}
g(\yvec)\,\mk(\xvec,\yvec)\,\diff S_{\yvec}.
\end{equation}
The main technical objective of this paper is to determine the boundary kernel
$\mk(\xvec,\yvec)$ explicitly in representative geometries and to relate it to
elliptic boundary-value problems \cite{jost2007partial} driven by the generator of the diffusion.

At this stage, the exit time $\tau_{\Omega}$ is treated as an internal variable
and is marginalized out in the analysis, unless otherwise stated.

\section{A unified framework for determining first-hitting location laws}
\label{sec:generator}
\begin{table*}[t]
\caption{Parabolic PDE versus generator-based routes to first-hitting location laws}
\centering
\begin{tabular}{@{}c p{6.5cm} @{\hspace{2em}} p{6.5cm}@{}}
\hline
& \textbf{Parabolic PDE route} & \textbf{Generator--Green-function route} \\ \hline
\textbf{Step 1} &
Construct a free-space fundamental solution for a parabolic partial differential equation (PDE). &
Pass to the infinitesimal generator of the Markov semigroup. \\ \hline
\textbf{Step 2} &
Impose absorbing boundary conditions (e.g., by image methods). &
Solve an elliptic boundary-value problem with Dirichlet data. \\ \hline
\textbf{Step 3} &
Compute boundary flux and integrate over all time. &
Extract the boundary kernel from the normal derivative of
the associated Green function. \\ \hline
\end{tabular}
\label{table:compare}
\end{table*}

This section presents a unified route, summarized in Table~\ref{table:compare}, for determining FHL laws induced by drift--diffusion processes in domains with absorbing boundaries. 
Related generator-based descriptions of boundary observables and boundary local time have been extensively studied in the context of reflected and absorbed diffusions; see, e.g., \cite{grebenkov2007boundary}.
Table~\ref{table:dictionary} records the corresponding stochastic and transport terminology used below.

We refer to the exit location on the boundary as the FHL. 
The approach relies on standard tools from stochastic analysis and potential theory, in particular, the infinitesimal generator of the diffusion semigroup and Dynkin’s formula \cite{doob2001classical,pinsky1995positive}.
By reformulating the problem at the level of the generator, the determination of FHL laws is reduced to properties of an elliptic boundary-value problem.

At a conceptual level, the procedure consists of three steps, summarized in Table~\ref{table:compare}.
The essential observation is that once the elliptic Dirichlet Green function
$\mathcal{G}(\xvec,\yvec)$  associated with the generator is available, the exit law on the boundary admits a kernel representation of the form \cite{doob2001classical,pinsky1995positive,grebenkov2007spectral}:
\begin{equation}
\mk(\xvec,\yvec)
=
-\,\partial_{n(\yvec)}\mathcal{G}(\xvec,\yvec),
\quad \yvec\in\partial\Omega,
\label{eq:main-preview}
\end{equation}
where $\partial_{n(\yvec)}$ denotes differentiation along the outward normal at the boundary point $\yvec$.

Equivalent time-dependent (parabolic) approaches recover the same exit law by
integrating the boundary flux over time \cite{redner2001guide,risken1989}.
The generator-based formulation instead begins with the time-marginalized
observable and obtains its boundary kernel directly from an elliptic problem.

We now derive the generator-based representation and specify the assumptions under which it holds.

\subsection{Infinitesimal generator and Markov semigroup}
Let $\{\mathbf{X}_t\}_{t\ge 0}$ be the It\^{o} diffusion introduced in
Sec.~\ref{sec:model}, evolving in $\Omega$ with absorbing boundary
$\partial\Omega$.
Throughout this work, we restrict attention to the constant-drift,
isotropic-diffusion setting specified in Sec.~\ref{sec:model}.

For a time-homogeneous Markov process, the operators
\begin{equation}
P_t f(\xvec)
:=
\mathbb{E}_{\xvec}\!\left[f(\mathbf{X}_t)\right],
\quad t\ge 0,
\end{equation}
define a Markov semigroup $\{P_t\}_{t\ge 0}$ acting on suitable test functions
$f$.
The infinitesimal generator $\mathcal{L}$ is defined on its domain
$\mathcal{D}(\mathcal{L})$ by
\begin{equation}
\mathcal{L}f
:=
\lim_{t\to 0^+}
\frac{P_t f - f}{t},
\quad f\in\mathcal{D}(\mathcal{L}).
\label{eq:def-generator}
\end{equation}

Applying It\^{o}'s formula to $f(\mathbf{X}_t)$ for $f\in C^2$ and using the
stochastic differential equation
\(
\diff \mathbf{X}_t
=
\drift\,\diff t
+
\sigma\,\diff \mathbf{B}_t
\),
we obtain
\begin{align}
\diff f(\mathbf{X}_t)
=\ &
(\nabla f)^\mathsf{T}\diff\mathbf{X}_t
+
\frac{1}{2}
(\diff\mathbf{X}_t)^\mathsf{T}
\mathrm{Hess}(f)
(\diff\mathbf{X}_t)
\\
=\ &
\Big[
\drift^\mathsf{T}\nabla f
+
\frac{\sigma^2}{2}\,\Delta f
\Big]\diff t
+
(\nabla f)^\mathsf{T}
\sigma\,\diff\mathbf{B}_t .
\end{align}
Taking expectations and using
$\mathbb{E}_{\xvec}[\diff\mathbf{B}_t]=0$, we arrive at the explicit form of the
generator
\begin{equation}
\mathcal{L}f(\xvec)
=
\drift^\mathsf{T}\nabla f(\xvec)
+
\frac{\sigma^2}{2}\,\Delta f(\xvec),
\label{eq:L-explicit}
\end{equation}
which is a second-order elliptic operator with constant coefficients.

We next consider the Dirichlet boundary-value problem
\begin{equation}
\left\{
\begin{array}{ll}
\mathcal{L}\phi=0 & \text{in }\Omega,\\
\phi=g & \text{on }\partial\Omega,
\end{array}
\right.
\label{eq:BVP1-1}
\end{equation}
where $g$ is a prescribed boundary function on $\partial\Omega$.

\subsection{Green-function representation of the exit measure}
\label{subsection:D}
From the definition of the generator,
\begin{equation}
\mathbb{E}_{\xvec}[f(\mathbf{X}_t)]
=
f(\xvec)
+
\mathbb{E}_{\xvec}
\!\left[
\int_0^t \mathcal{L}f(\mathbf{X}_s)\,\diff s
\right],
\label{generator-int}
\end{equation}
for deterministic time $t\in\R^+$.
Dynkin’s formula extends this identity to stopping times.
Let $\tau_\Omega$ denote the exit time from $\Omega$, and assume
$\mathbb{E}_{\xvec}[\tau_\Omega]<\infty$.
Then for $f\in C^2$,
\begin{equation}
\mathbb{E}_{\xvec}[f(\mathbf{X}_{\tau_\Omega})]
=
f(\xvec)
+
\mathbb{E}_{\xvec}
\!\left[
\int_0^{\tau_\Omega}
\mathcal{L}f(\mathbf{X}_s)\,\diff s
\right].
\label{formula-dynkin}
\end{equation}

If $\phi$ solves the Dirichlet problem~\eqref{eq:BVP1-1}, then
$\mathcal{L}\phi=0$ in $\Omega$.
Setting $f=\phi$ in~\eqref{formula-dynkin} yields
\begin{equation}
\phi(\xvec)
=
\mathbb{E}_{\xvec}[g(\mathbf{X}_{\tau_\Omega})].
\label{eq:phi-exit}
\end{equation}
Introducing the exit measure $\omega^{\xvec}$ on $\partial\Omega$, this relation
can be written as
\begin{equation}
\phi(\xvec)
=
\int_{\partial\Omega}
g(\yvec)\,\omega^{\xvec}(\diff\yvec).
\label{eq:exit-measure}
\end{equation}

On the other hand, solutions of elliptic boundary-value problems admit Green
function representations.
Let $\mathcal{G}(\xvec,\yvec)$ denote the Dirichlet Green function of the
elliptic operator $\mathcal{L}$.
Then
\begin{equation}
\phi(\xvec)
=
\int_{\partial\Omega}
g(\yvec)\,
\bigl[-\partial_{n(\yvec)}\mathcal{G}(\xvec,\yvec)\bigr]\,
\diff S_{\yvec}.
\label{general-green-2}
\end{equation}
Comparing~\eqref{eq:exit-measure} and~\eqref{general-green-2} yields the
representation
\begin{equation}
\omega^{\xvec}(\diff\yvec)
=
\mk(\xvec,\yvec)\,\diff S_{\yvec},
\quad
\mk(\xvec,\yvec)
=
-\,\partial_{n(\yvec)}\mathcal{G}(\xvec,\yvec),
\label{eq:main}
\end{equation}
which expresses the exit law on $\partial\Omega$ as a boundary kernel given by
the outward normal derivative of the elliptic Green function.
Explicit evaluations of $\mk(\xvec,\yvec)$ for representative geometries are given
in Sec.~\ref{sec:case}.

\section{Calculating boundary kernels in representative geometries}
\label{sec:case}

\begin{figure}[!t]
\centering
\resizebox{0.48\textwidth}{!}{%
\begin{tikzpicture}[
    >=Latex,
    node distance=2cm,
    font=\sffamily\small,
    line cap=round,
    line join=round
]
    \def\BxX{0}      
    \def\IxX{5}      
    \def\Ymax{2.5}
    \def\Ymin{-2.5}

    \colorlet{rxcol}{black!85}
    \colorlet{txcol}{black!65}
    \colorlet{pathcol}{teal!70!black}
    \colorlet{driftcol}{purple!70!black}
    \colorlet{boxfill}{gray!10}

    \draw[rxcol, line width=1.1pt] (\BxX, \Ymin) -- (\BxX, \Ymax)
        node[above] {Absorbing boundary};
    \fill[pattern=horizontal lines, pattern color=black!25]
        (\BxX-0.3, \Ymin) rectangle (\BxX, \Ymax);

    \draw[txcol, line width=1.0pt, dotted] (\IxX, \Ymin) -- (\IxX, \Ymax)
        node[above] {Initial reference plane};

    \node[draw=black!70, fill=boxfill, rounded corners=2pt,
          minimum width=2cm, minimum height=1cm, align=center] (sink)
          at (\BxX-2.5, 0) {First-Hitting\\Location};
    \draw[->, line width=0.9pt] (\BxX, 0) -- (sink);

    \node[draw=black!70, fill=boxfill, rounded corners=2pt,
          minimum width=2cm, minimum height=1cm, align=center] (source)
          at (\IxX+2.5, 0) {Initial\\Position};
    \draw[->, line width=0.9pt] (source) -- (\IxX, 0);

    \fill[pathcol] (\IxX, 0.5) circle (3pt);
    \fill[pathcol] (\BxX, 1.2) circle (3pt);

    \pgfmathsetseed{12471}
    \draw[pathcol, line width=1.0pt, decorate,
          decoration={random steps,segment length=3pt,amplitude=3.5pt}]
        (\IxX, 0.5) -- (4.0, 0.8) -- (2.7, 0.2) -- (1.3, 1.5) -- (\BxX, 1.2);

    \node[pathcol, font=\small, align=center]
    at (2.5, 1.9)
    {Stochastic Trajectory\\(sample path)};

    \begin{scope}[shift={(2.3, -2.35)}]
    \draw[->, thin, gray] (0,0) -- (0.5, 0) node[right] {\footnotesize $x_d$};
    \draw[->, thin, gray] (0,0) -- (0, 0.5) node[above] {\footnotesize $x_{\parallel}$};
    \end{scope}

    \draw[->, line width=0.95pt, color=driftcol] (3.8, -0.5) -- (1.8, -0.5)
    node[midway, below, font=\small, align=center]
    {Drift\\ (directed transport)};

    \draw[<->, line width=0.9pt] (\BxX, \Ymin-0.3) -- (\IxX, \Ymin-0.3)
        node[midway, below] {$\lambda$};

    \draw[black!25, thin, dashed] (\BxX, \Ymin) -- (\BxX, \Ymin-0.5);
    \draw[black!25, thin, dashed] (\IxX, \Ymin) -- (\IxX, \Ymin-0.5);

\end{tikzpicture}%
}
\caption{
    Schematic illustration of an FHL process in an absorbing half-space.
    A drift--diffusion trajectory is initialized at a normal distance $\lambda$
    from the absorbing boundary and evolves in the bulk until it is terminated
    upon first contact with the boundary.
    The spatial location of this exit event defines the primary observable of interest.
    The figure provides an interpretive visualization of how stochastic transport
    maps initial conditions to boundary exit measures.
    }
\label{fig:channel_model_d2}
\end{figure}

This section applies the general exit-law representation~\eqref{eq:main}
to geometries in which the boundary kernel $\mk(\xvec,\yvec)$ can be evaluated in
closed form \cite{redner2001guide,doob2001classical,grebenkov2007spectral}.
We focus on the absorbing half-space, which provides the simplest extended
boundary geometry and serves as a canonical local model for smooth interfaces in
the limit of vanishing curvature.

\begin{figure*}[!t]
    \centering
    \includegraphics[width=0.8\textwidth]{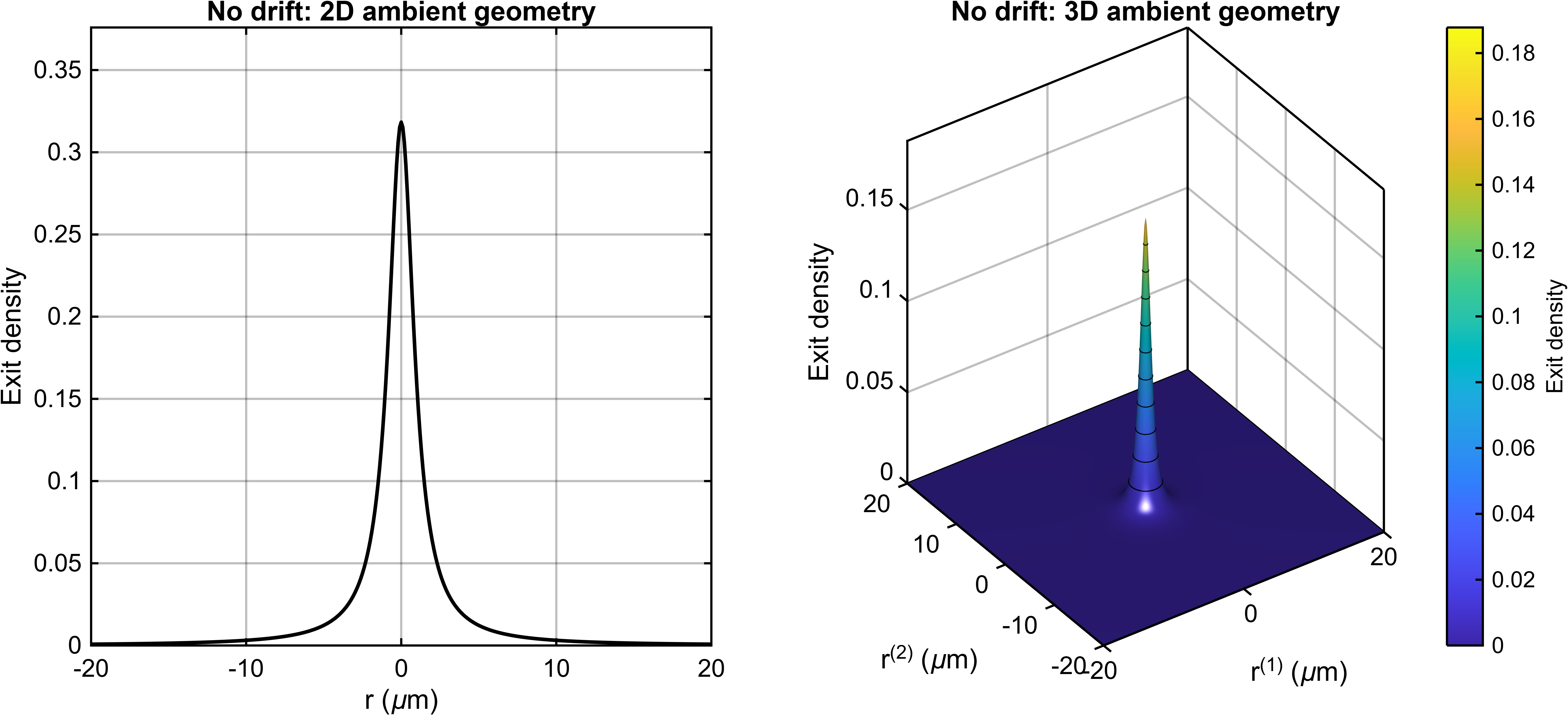}
    \caption{Boundary hitting distributions in the drift-free regime ($\boldsymbol{u}=0$),
    shown for two- and three-dimensional ambient geometries.
    In the absence of drift, the first-hitting location law is isotropic and
    exhibits scale-free algebraic decay, corresponding to a Cauchy-type boundary
    measure \cite{doob2001classical}
 without an intrinsic length scale.}
    \label{fig:no_drift}
\end{figure*}

\begin{figure*}[!t]
    \centering
    \includegraphics[width=0.8\textwidth]{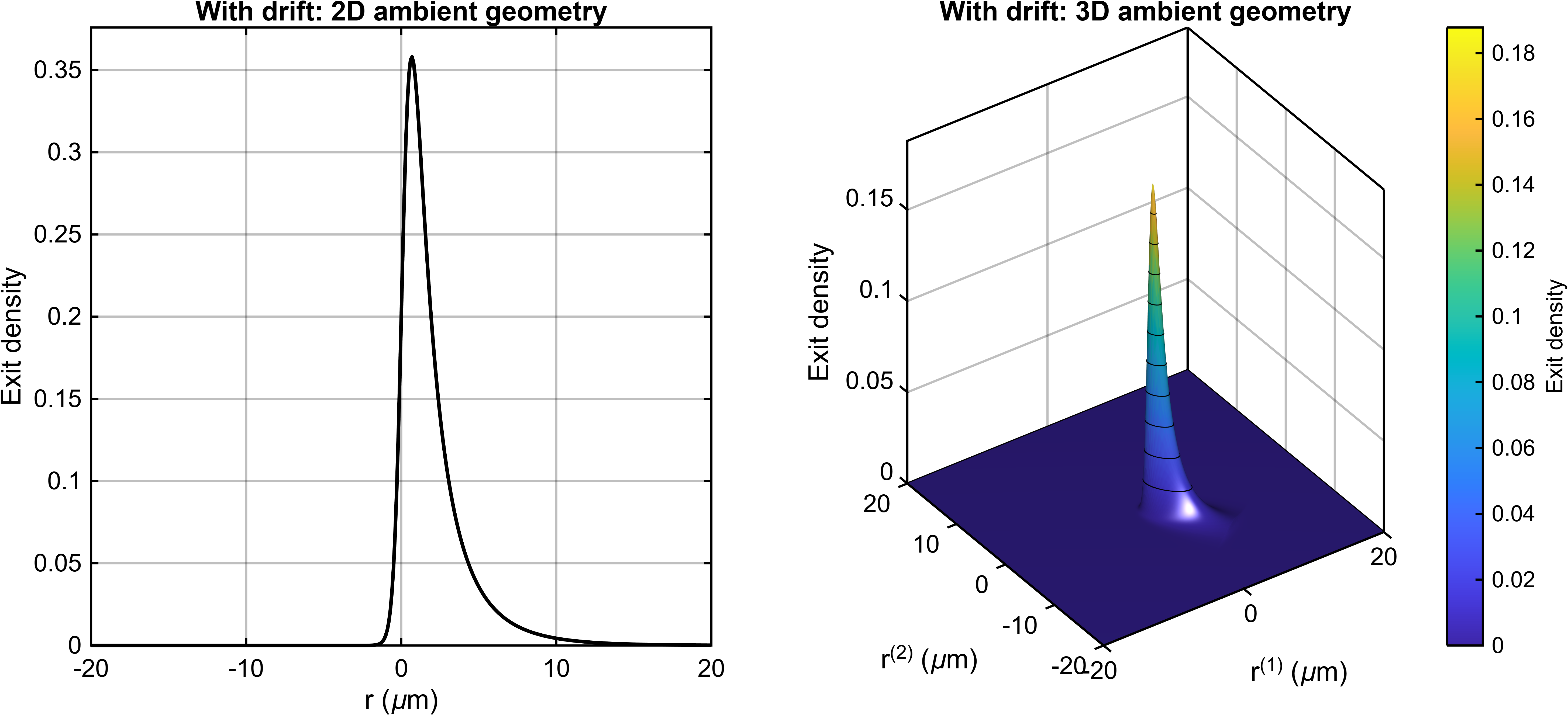}
    \caption{Boundary hitting distributions in the presence of drift
    ($\boldsymbol{u}\neq 0$), shown for two- and three-dimensional ambient geometries.
    Drift introduces an intrinsic length scale and exponentially suppresses
    large boundary displacements, leading to localized and anisotropic hitting
    profiles consistent with the drift-regularized analytical kernels.}
    \label{fig:drifted}
\end{figure*}

Within this setting, two cases are of primary interest: (i)~a two-dimensional system with an absorbing line (Fig.~\ref{fig:channel_model_d2}) and (ii)~a three-dimensional system with an absorbing plane (Fig.~\ref{fig:channel_model_d3}).
Both cases admit explicit boundary kernels under constant drift and isotropic
diffusion, yielding analytically tractable expressions that make clear how
directed transport regularizes diffusion-induced spatial fluctuations.
These hyperplanar results constitute one of the main explicit contributions of the present
work and serve as benchmark solutions for drift--diffusion exit statistics.
Representative boundary-hitting profiles in the drift-free and drifted regimes
are shown in Figs.~\ref{fig:no_drift} and~\ref{fig:drifted}, respectively.

\begin{figure}[!t]
    \centering
    \resizebox{\linewidth}{!}{%
    \begin{tikzpicture}[
        x={(1.4cm,0cm)},
        y={(0.5cm,0.4cm)},
        z={(0cm,1cm)},
        scale=2.0,
        >=stealth
    ]

    \def\hitY{0.6}
    \def\hitZ{0.5}

    \colorlet{rxcol}{black!85}
    \colorlet{txcol}{black!65}
    \colorlet{pathcol}{teal!70!black}
    \colorlet{driftcol}{purple!70!black}
    \colorlet{boxfill}{gray!10}

    \fill[blue!5, opacity=0.6]
        (0,-1,-1) -- (0,1,-1) -- (0,1,1) -- (0,-1,1) -- cycle;
    \draw[blue!30]
        (0,-1,-1) -- (0,1,-1) -- (0,1,1) -- (0,-1,1) -- cycle;

    \node[anchor=north east, xshift=-3pt] at (0,-1,-1)
        {\footnotesize \textbf{Absorbing boundary} $\partial\Omega$ ($x_d=0$)};

    \fill[gray!10, opacity=0.7]
        (1,-1,-1) -- (1,1,-1) -- (1,1,1) -- (1,-1,1) -- cycle;
    \draw[gray!40, dashed]
        (1,-1,-1) -- (1,1,-1) -- (1,1,1) -- (1,-1,1) -- cycle;

    \node[anchor=north west, xshift=3pt] at (1,-1,-1)
        {\footnotesize \textbf{Initial plane} ($x_d=\lambda$)};

    \node[anchor=west] at (1,0,0)
        {\footnotesize $\xvec_0$};
    \fill[black] (1,0,0) circle (1.5pt);

    \draw[->, thick, color=driftcol]
        (0.85,0.1,0.2) -- ++(-0.35,0.15,0.1)
        node[midway, above, sloped]
        {\footnotesize Drift $\drift$};

    \pgfmathsetseed{12473}
    \draw[
        color=pathcol,
        thick,
        decorate,
        decoration={
            random steps,
            segment length=3pt,
            amplitude=2pt
        }
    ]
        (1,0,0) -- (0.55,0.3,0.2) -- (0,\hitY,\hitZ);

    \draw[->, black]
        (0.05,-1.4,-1.2) -- (0.95,-1.4,-1.2)
        node[midway, below]
        {\footnotesize Normal distance $\lambda$};

    \draw
        (0.05,-1.3,-1.2) -- (0.05,-1.5,-1.2);
    \draw
        (0.95,-1.3,-1.2) -- (0.95,-1.5,-1.2);

    \draw[->, thin, gray]
        (-0.9,0,0) -- (-0.6,0,0)
        node[right] {\scriptsize $x_d$};

    \draw[->, thin, gray]
        (-0.9,0,0) -- (-0.9,0.4,0)
        node[right] {\scriptsize $x_{\parallel}^{(1)}$};

    \draw[->, thin, gray]
        (-0.9,0,0) -- (-0.9,0,0.3)
        node[above] {\scriptsize $x_{\parallel}^{(2)}$};

    \coordinate (Hit) at (0,\hitY,\hitZ);
    \fill[red] (Hit) circle (1.5pt);

    \node[
        anchor=south,
        yshift=6pt,
        align=center,
        font=\footnotesize
    ] at (Hit) {
        \textbf{First-Hitting Location} $\mathbf{X}_{\tau_\Omega}$\\
        at time $\tau_\Omega$
    };

    \end{tikzpicture}%
    }
\caption{
    Schematic illustration of a drift--diffusion process in a half-space
    with an absorbing boundary (shown for $d=3$).
    A trajectory initialized at $\xvec_0\in\Omega$, at normal distance
    $\lambda$ from the absorbing boundary $\partial\Omega$, evolves in
    the bulk under a constant drift $\drift$ and isotropic diffusion,
    and terminates upon first reaching $\partial\Omega$.
    The corresponding exit time $\tau_\Omega$ and boundary location
    $\mathbf{X}_{\tau_\Omega}$ define the induced boundary law.
    The figure highlights the geometric setting and variables underlying
    the first-hitting location framework analyzed in the text.
}
\label{fig:channel_model_d3}
\end{figure}

\subsection{Absorbing half-space: reduction to an elliptic problem}
We specialize the stochastic transport model defined in Eq.~\eqref{eq:IP} to the case of an absorbing half-space $\Omega$ with boundary $\partial\Omega$. 
Under the constant-drift and isotropic-diffusion assumptions of Eq.~\eqref{ad-eq}, 
the infinitesimal generator $\mathcal{L}$ of the process is a second-order elliptic operator with drift:
\begin{equation}
\mathcal{L}
=
\drift^{\mathsf T} \nabla
+
\frac{\sigma^2}{2} \Delta .
\end{equation}
The boundary kernel $\mk(\xvec,\yvec)$ defined in Eq.~\eqref{eq:fhl-def} is obtained from the outward normal derivative of the Green function associated with the Dirichlet problem $\mathcal{L}\phi=0$ in $\Omega$.

To determine $\mk(\xvec,\yvec)$, we reduce the problem to a Helmholtz form by exploiting the scaled drift vector $\boldsymbol{u}$ introduced in Sec.~\ref{sec:model}. 
Applying the exponential change of variables $\psi(\xvec) = \exp(\boldsymbol{u}^{\mathsf T}\xvec) \phi(\xvec)$, the Dirichlet problem transforms into a Helmholtz-type equation for $\psi$:
\begin{equation}
\left\{
\begin{array}{ll}
(\Delta-\|\boldsymbol{u}\|^2)\psi=0 & \text{in }\Omega,\\
\psi=\tilde g & \text{on }\partial\Omega,
\end{array}
\right.
\label{eq:BVP-Helm_u}
\end{equation}
where $\tilde g(\yvec) = e^{\boldsymbol{u}^{\mathsf T}\yvec}g(\yvec)$ \cite{risken1989,doob2001classical,pinsky1995positive}. 
By adopting the normalized parameter $\boldsymbol{u}$, the diffusion strength $\sigma^2$ is implicitly absorbed into the geometric scale. 
Let $\tilde{\mathcal G}(\xvec,\yvec)$ denote the Green function associated with \eqref{eq:BVP-Helm_u}. 
The boundary kernel $\mk(\xvec,\yvec)$ then admits the following exact representation:
\begin{equation}
\mk(\xvec,\yvec)
=
e^{\boldsymbol{u}^{\mathsf T}(\yvec-\xvec)}
\bigl[-\partial_{n(\yvec)}\tilde{\mathcal G}(\xvec,\yvec)\bigr],
\quad \yvec\in\partial\Omega .
\label{eq:FHL-kernel-from-Helm_final}
\end{equation}

\subsection{Two-dimensional case: absorbing line}
\label{subsec:2D-line}
For $d=2$, the Helmholtz Green function in the half-space admits the image
representation \cite{polyanin2001handbook}:
\begin{equation}
\tilde{\mathcal G}(\xvec,\yvec)
=
\frac{1}{2\pi}\!\left[
K_0(\|\boldsymbol{u}\|\mathcal{R}_1)-K_0(\|\boldsymbol{u}\|\mathcal{R}_2)
\right],
\end{equation}
with
\begin{subequations}\label{eq:rho-def}
\begin{align}
\mathcal{R}_1
&=\sqrt{(x_1-\xi)^2+(x_2-\eta)^2},
\\
\mathcal{R}_2
&=\sqrt{(x_1-\xi)^2+(x_2+\eta)^2},
\end{align}
\end{subequations}
where $\mathcal{R}_2$ corresponds to the standard image point associated with the
absorbing boundary.

Fix $\xvec=(x_1,\lambda)$ and $\yvec=(\xi,0)$, where $\lambda>0$ denotes the
normal distance from the initial point to the boundary.
Differentiation in the outward normal direction yields
\begin{equation}
-\partial_{n(\yvec)}\tilde{\mathcal G}(\xvec,\yvec)
=
\frac{\|\boldsymbol{u}\|\lambda}{\pi}
\frac{K_1(\|\boldsymbol{u}\|\rho)}{\rho},~\rho=\sqrt{(x_1-\xi)^2+\lambda^2}.
\end{equation}

Substituting this expression into the exponential tilting
representation~\eqref{eq:FHL-kernel-from-Helm_final}, the boundary kernel along
the absorbing line is given explicitly by
\begin{equation}
\label{eq:2Dresult-PRE}
\mk(\xvec,\yvec)
=
\frac{\|\boldsymbol{u}\|\lambda}{\pi}
\exp\!\bigl\{\boldsymbol{u}^{\mathsf T}(\yvec-\xvec)\bigr\}
\frac{K_1(\|\boldsymbol{u}\|\rho)}{\rho},
\quad
\yvec\in\partial\Omega .
\end{equation}

\subsection{Three-dimensional case: absorbing plane}
\label{subsec:3D-plane}
For $d=3$, the Helmholtz Green function in the half-space is given by \cite{polyanin2001handbook}: 
\begin{equation}
\tilde{\mathcal G}(\xvec,\yvec)
=
\frac{e^{-\|\boldsymbol{u}\|\mathcal R_1}}{4\pi\mathcal R_1}
-
\frac{e^{-\|\boldsymbol{u}\|\mathcal R_2}}{4\pi\mathcal R_2},
\end{equation}
with
\begin{subequations}\label{eq:Rdef}
\begin{align}
\mathcal{R}_1
&=\sqrt{(x_1-\xi)^2+(x_2-\eta)^2+(x_3-\zeta)^2},
\\
\mathcal{R}_2
&=\sqrt{(x_1-\xi)^2+(x_2-\eta)^2+(x_3+\zeta)^2},
\end{align}
\end{subequations}
where $\mathcal{R}_2$ corresponds to the standard image construction associated
with the absorbing boundary.

Fix $\xvec=(x_1,x_2,\lambda)$ and $\yvec=(\xi,\eta,0)$, where $\lambda>0$ denotes
the normal distance from the initial point to the boundary.
After taking the outward normal derivative and applying the exponential tilting
relation~\eqref{eq:FHL-kernel-from-Helm_final}, the boundary kernel on the
absorbing plane takes the explicit form
\begin{equation}
\label{eq:3Dresult-PRE}
\mk(\xvec,\yvec)
=
\frac{\lambda}{2\pi}
\exp\!\bigl\{\boldsymbol{u}^{\mathsf T}(\yvec-\xvec)\bigr\}
e^{-\|\boldsymbol{u}\|\rho}
\frac{1+\|\boldsymbol{u}\|\rho}{\rho^3},
\end{equation}
where $\rho=\|\yvec-\xvec\|$.

\subsection{\texorpdfstring{General $d$-dimensional exit law}{General d-dimensional exit law}}
\label{subsec:higher_dim}
The explicit results in two and three dimensions reveal a
common, dimension-independent structure of boundary exit laws in the absorbing
half-space.
In the unified convention used throughout the paper, the ambient space has dimension $d$, while the induced boundary kernel lives on a $(d-1)$-dimensional absorbing interface.
Guided by low-dimensional expressions, we are naturally led to the following general expression
for the boundary kernel in $\mathbb{R}^{d}$:
\begin{align}
\label{eq:dD-FHL}
\mathcal{K}^{(d-1)}(\boldsymbol{r};\boldsymbol{u},\lambda)
&=
2\lambda
\frac{\|\boldsymbol{u}\|^{\frac{d}{2}}}{(2\pi)^{\frac{d}{2}}}
\exp\!\big(
\boldsymbol{u}_{\parallel}^{\mathsf T}\boldsymbol{r}
-
u_d\lambda
\big)
\nonumber\\
&\quad\times
\frac{
K_{\frac{d}{2}}
\!\left(
\|\boldsymbol{u}\|
\sqrt{\|\boldsymbol{r}\|^2+\lambda^2}
\right)
}{
\left(
\sqrt{\|\boldsymbol{r}\|^2+\lambda^2}
\right)^{\frac{d}{2}}
}.
\end{align}
Here $\boldsymbol{r}\in\mathbb{R}^{d-1}$ denotes the tangential displacement along
the absorbing boundary.

A concise derivation of Eq.~\eqref{eq:dD-FHL} based on the heat-kernel resolvent
representation of the modified Helmholtz operator is provided in
Appendix~\ref{app:proof_dDhk}.
The structure of the kernel is strongly constrained by symmetry and scaling considerations.
In particular, translational invariance along the absorbing boundary,
rotational symmetry in the $d-1$ boundary-parallel directions,
and the exponential tilting induced by the drift together single out the modified
Bessel kernel appearing above.
From this perspective, Eq.~\eqref{eq:dD-FHL} should be viewed as the natural extension
of the exact two- and three-dimensional results to general ambient dimension.

As a sanity check, we specialize the general expression~\eqref{eq:dD-FHL}
to the case $d=2$, corresponding to a two-dimensional ambient space with a
one-dimensional absorbing boundary.
In this setting, the tangential displacement reduces to a scalar
$r\in\mathbb{R}$, and the modified Bessel function appearing in
\eqref{eq:dD-FHL} becomes $K_1$ \cite{Abramowitz1972,pinsky1995positive}.
Substituting $d=2$ and simplifying the prefactors, the general formula reduces
exactly to
\begin{equation*}
\mathcal{K}^{(1)}(r;\boldsymbol{u},\lambda)
=
\frac{\|\boldsymbol{u}\|\lambda}{\pi}
\exp\!\big(u_\parallel r-u_d\lambda\big)
\frac{K_1\!\left(\|\boldsymbol{u}\|\sqrt{r^2+\lambda^2}\right)}
{\sqrt{r^2+\lambda^2}},
\end{equation*}
which coincides with the explicit boundary kernel obtained for the absorbing
line in Sec.~\ref{subsec:2D-line}, cf.~\eqref{eq:2Dresult-PRE}.

A further consistency check is obtained by specializing
\eqref{eq:dD-FHL} to the case $d=3$, corresponding to a three-dimensional
ambient space with a two-dimensional absorbing boundary.
In this case, the modified Bessel function reduces to the half-integer order
$K_{3/2}$, which admits a closed-form representation.
Substituting $d=3$ into \eqref{eq:dD-FHL} and simplifying the resulting
expression using the identity for $K_{3/2}$ yields
\begin{equation}
\label{eq:kernel_slim}
\mathcal{K}^{(2)}(\boldsymbol{r};\boldsymbol{u},\lambda)
=
\frac{\lambda e^{\boldsymbol{u}_\parallel^{\mathsf T}\boldsymbol{r}-u_d\lambda}}{2\pi}
\cdot
\frac{e^{-\|\boldsymbol{u}\|\rho}(1+\|\boldsymbol{u}\|\rho)}{\rho^3},
\end{equation}
where $\rho = \sqrt{\|\boldsymbol{r}\|^2+\lambda^2}$ denotes the slant distance.
The above form coincides exactly with the boundary kernel obtained for the absorbing
plane in Sec.~\ref{subsec:3D-plane}, cf.~\eqref{eq:3Dresult-PRE}.
Together with the $d=2$ case, this agreement confirms that Eq.~\eqref{eq:dD-FHL}
correctly reproduces the known closed-form results in both two and three dimensions.

\begin{figure*}[t]
\centering
\includegraphics[width=0.95\textwidth]{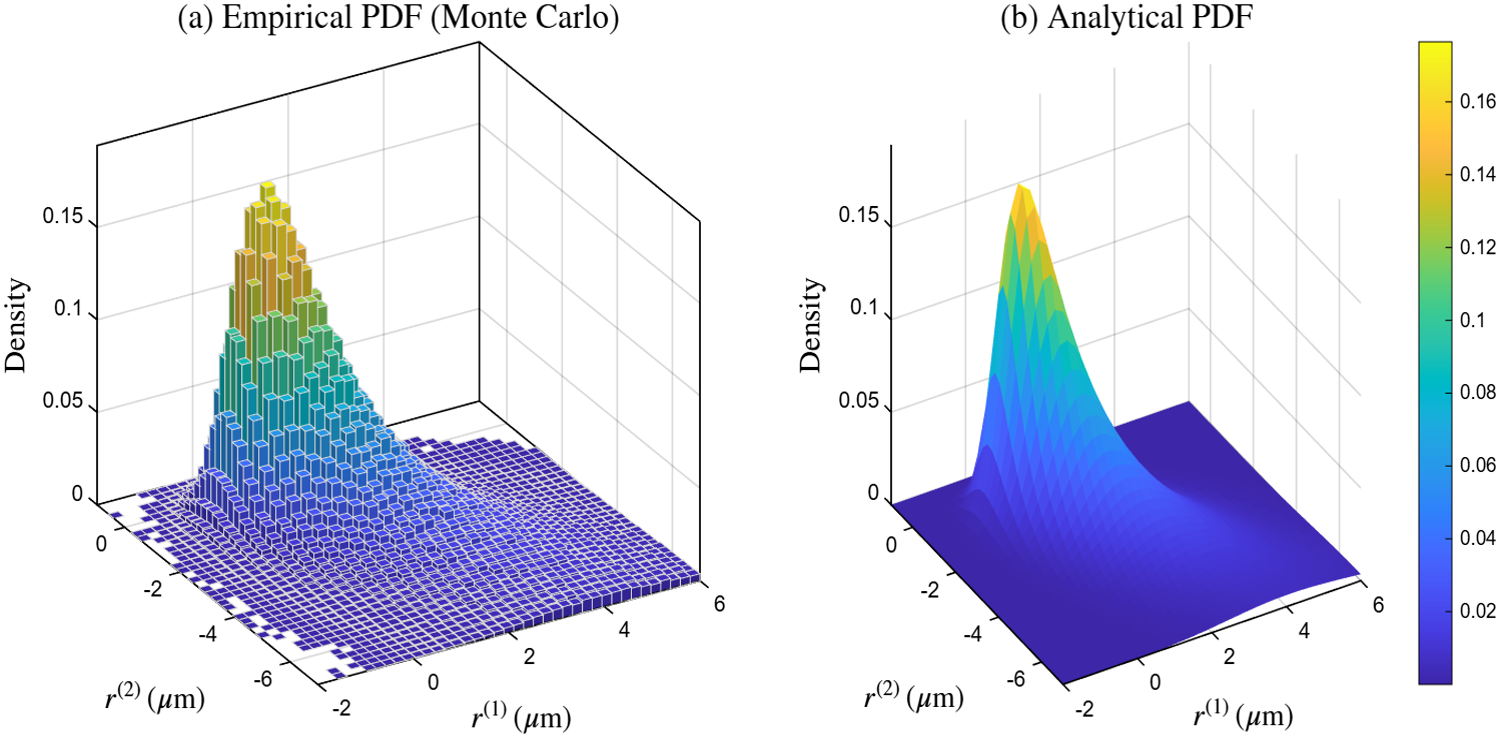}
\caption{Numerical validation of the FHL distribution on a planar absorbing boundary in a three-dimensional ambient geometry. (a) Empirical PDF obtained from $10^6$ Monte Carlo trajectories of the Langevin dynamics Eq.~\eqref{eq:IP}. (b) Analytical boundary kernel calculated via Eq.~\eqref{eq:kernel_slim}. The parameters are set to $\lambda = 1.0\,\mu\text{m}$, $\sigma = 1.0\,\mu\text{m}/\sqrt{\text{s}}$, and a constant drift $\mathbf{v} = [2, -3, -1]\,\mu\text{m/s}$. The simulation results are in excellent agreement with the analytical prediction, accurately capturing the drift-induced anisotropy and the localized hitting profile.}
\label{fig:numerical_validation}
\end{figure*}

\subsection{Numerical validation via Monte Carlo simulation}
To validate the analytical boundary kernels derived in this section, we perform particle-based Monte Carlo simulations of the underlying Langevin dynamics Eq.~\eqref{eq:IP}. We consider a 3D system with an absorbing plane at $x_3 = 0$, an initial normal distance $\lambda = 1.0\,\mu\text{m}$, and a noise strength $\sigma = 1.0\,\mu\text{m}/\sqrt{\text{s}}$. A uniform drift $\mathbf{v} = [2, -3, -1]\,\mu\text{m/s}$ is applied to introduce both normal and tangential transport components.

As shown in Fig.~\ref{fig:numerical_validation}, the spatial distribution of the first-hitting locations obtained from $10^6$ independent trajectories is in excellent agreement with the analytical prediction of Eq.~\eqref{eq:3Dresult-PRE}. Specifically, Fig.~\ref{fig:numerical_validation}(a) displays the empirical probability density function (PDF) sampled from the simulated hitting events, while Fig.~\ref{fig:numerical_validation}(b) shows the corresponding theoretical kernel. In particular, the theory accurately captures the drift-induced anisotropy and the exponential suppression of tangential excursions along the planar interface. This agreement confirms that the generator-based framework provides a robust structural description of the microscopic stochastic transport process, effectively bridging the pathwise Langevin dynamics with the macroscopic boundary exit law.

\section{Asymptotic structure and geometric diagnostics of exit laws}
\label{sec:information}
The planar results of Sec.~\ref{sec:case} provide explicit \emph{exit laws} on the
absorbing boundary, arising solely from the interplay between stochastic dynamics
and geometry.
Recalling the kernel representation established in Sec.~\ref{sec:generator},
the boundary exit measure $\omega^{\xvec}(\diff\yvec)$ can be written as
\begin{equation}
\omega^{\xvec}(\diff\yvec)
=
\mathcal{K}(\xvec,\yvec)\,\diff S_{\yvec},
\qquad
\yvec\in\partial\Omega,
\end{equation}
where $\mathcal{K}(\xvec,\yvec)$ denotes the boundary kernel and $\diff S_{\yvec}$ is the
surface measure on $\partial\Omega$.
These explicit kernel representations allow us to go beyond a purely structural
description of exit laws and to examine their asymptotic organization across
diffusion-dominated and drift-regularized regimes.

Although differential entropy is an information-theoretic
functional, here it is used only to construct a compact geometric diagnostic of
the spatial spread of first-hitting locations. We do not treat this diagnostic
as an independent information-theoretic development; a full input--output
analysis would additionally require a specified input ensemble, modulation, and
receiver model. From a statistical-physics perspective, the resulting effective
width summarizes a geometric manifestation of nonequilibrium first-passage
dynamics rather than a communication protocol
\cite{Akyildiz2008,farsad2016,redner2001guide,bray2013persistence}.

\begin{figure*}[t]
\centering
\includegraphics[width=0.95\textwidth]{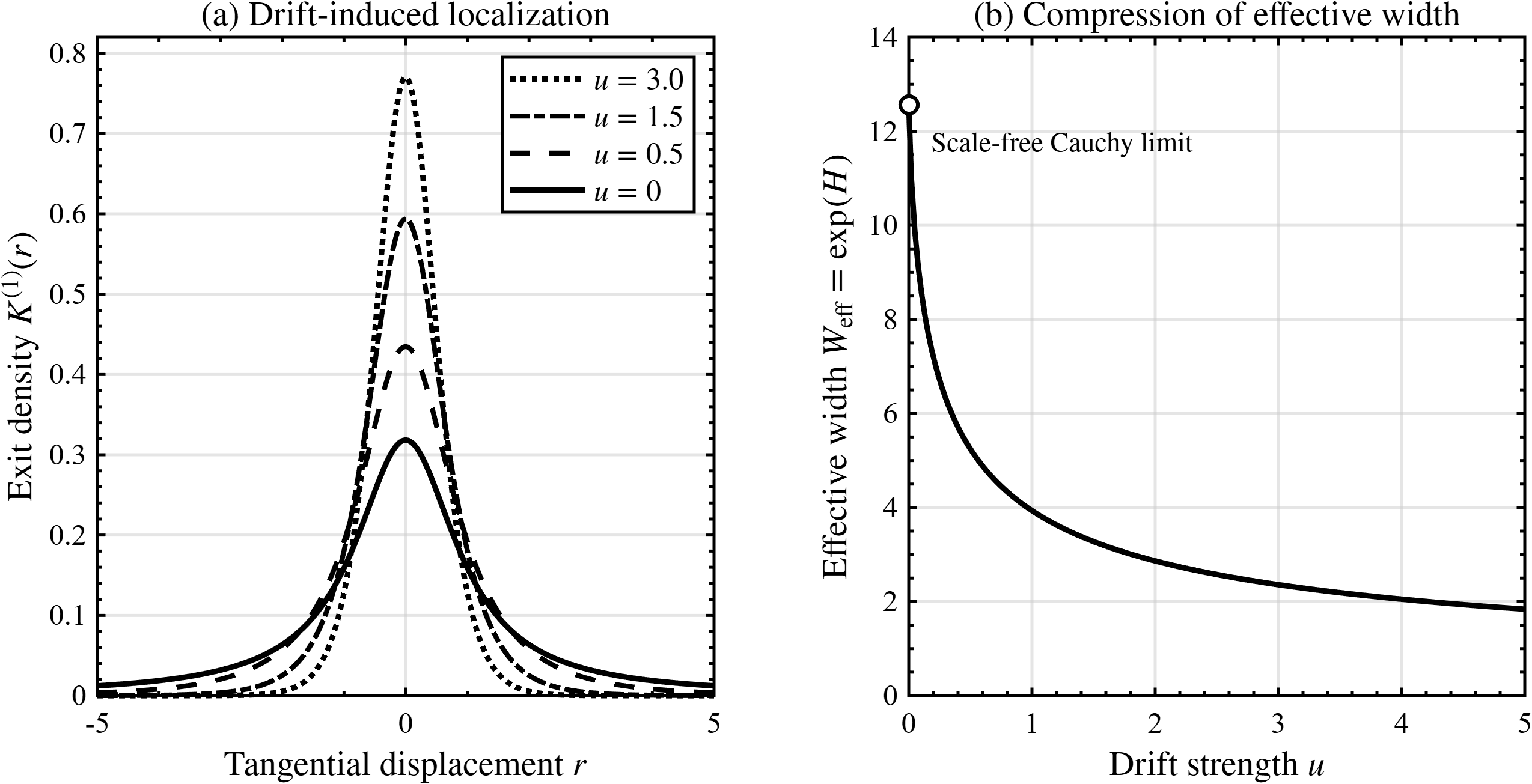}
\caption{Spatial localization and effective width compression. (a) Exit density $K^{(1)}(r)$ for varying drift strengths $u$, showing the transition from scale-free Cauchy decay ($u=0$) to exponential screening. (b) Effective width $W_{\text{eff}} = \exp(H)$ as a function of drift strength $u$. The sharp decrease quantifies the drift-induced regularization of boundary fluctuations. Unlike the variance, which diverges in the Cauchy limit (circle marker), $W_{\text{eff}}$ provides a robust and finite geometric measure of the spatial spread across all transport regimes.}
\label{fig:FHL_localization}
\end{figure*}

\subsection{Asymptotic structure of planar exit laws}
The explicit (hyper)-planar kernels derived in Sec.~\ref{sec:case} reveal a robust
asymptotic structure of boundary exit laws that is not tied to a specific spatial
dimension.
At large tangential displacements along the absorbing boundary, the qualitative
behavior of the exit law is governed primarily by the presence or absence of an
intrinsic length scale, rather than by microscopic details of the stochastic
transport mechanism.

In the drift-free case, the exit law coincides with harmonic measure
\cite{doob2001classical,pinsky1995positive,makarov1985distortion}.
The induced boundary kernel is scale-free and exhibits algebraic decay at large
distances.
For a $(d-1)$-dimensional absorbing boundary, the asymptotic tail takes the form
\begin{equation}
\label{eq:general_tail_nodrift}
\mathcal{K}^{(d-1)}(\boldsymbol{r})
\sim
\|\boldsymbol{r}\|^{-d},
\qquad
\|\boldsymbol{r}\|\to\infty,
\quad (u=0),
\end{equation}
where $\boldsymbol{r}\in\mathbb{R}^{d-1}$ denotes the boundary-parallel displacement
and $u=\|\boldsymbol{u}\|$ is the magnitude of the scaled drift vector.
This heavy-tailed behavior reflects the absence of any intrinsic spatial scale and
is a generic feature of boundary laws induced by harmonic kernels
\cite{redner2001guide,bray2013persistence}.
As a consequence, higher-order moments diverge, and moment-based or
variance-controlled descriptions fail, reflecting the intrinsically scale-free
nature of the exit statistics.
This asymptotic form follows directly from the general kernel representation
\eqref{eq:dD-FHL}: setting $\nu=d/2$ and
$\rho=\sqrt{\|\boldsymbol{r}\|^2+\lambda^2}$, the modified Bessel function appearing
in~\eqref{eq:dD-FHL} has argument $z=\|\boldsymbol{u}\|\,\rho$, so that in the
zero-drift limit $\|\boldsymbol{u}\|\to 0$ one has $z\to 0^+$ and the small-argument
expansion $K_\nu(z)\sim \tfrac12\Gamma(\nu)(2/z)^\nu$ \cite{Abramowitz1972}.
The prefactor $\|\boldsymbol{u}\|^\nu$ in~\eqref{eq:dD-FHL} cancels the divergence of
$K_\nu(z)$, yielding
\begin{equation}
\mathcal{K}^{(d-1)}(\boldsymbol{r};\boldsymbol{0},\lambda)
\propto
(\|\boldsymbol{r}\|^2+\lambda^2)^{-\nu},
\end{equation}
and hence $\mathcal{K}^{(d-1)}(\boldsymbol{r};\boldsymbol{0},\lambda)
\sim \|\boldsymbol{r}\|^{-2\nu}=\|\boldsymbol{r}\|^{-d}$ as
$\|\boldsymbol{r}\|\to\infty$.

When a nonzero drift component toward the absorbing boundary is present, an
intrinsic length scale $\ell_u\sim u^{-1}$ emerges and qualitatively alters the
asymptotic behavior.
For purely normal drift toward the absorbing boundary, the boundary kernel acquires an exponential
screening factor and obeys
\begin{equation}
\label{eq:general_tail_drift}
\begin{aligned}
\mathcal{K}^{(d-1)}(\boldsymbol{r})
&\sim
\exp\bigl(-u\,\|\boldsymbol{r}\|\bigr)\,
\|\boldsymbol{r}\|^{-(d+1)/2},\\
&\hspace{3em}\|\boldsymbol{r}\|\to\infty,\qquad u\neq 0 .
\end{aligned}
\end{equation}
This exponential screening effect follows from the large-argument behavior of the
modified Bessel function appearing in the planar kernel representation: for
$\|\boldsymbol{r}\|\to\infty$ at fixed $u>0$, the Bessel argument grows linearly
with $\|\boldsymbol{r}\|$, and the leading asymptotic form
$K_\nu(z)\sim \sqrt{\pi/(2z)}\,e^{-z}$ yields both the exponential cutoff
and the reduced algebraic prefactor.
Together with the drift-free case, these two regimes are governed respectively
by the small- and large-argument limits of the same modified Bessel function; the
relevant asymptotic expansions are summarized in Appendix~\ref{app:bessel}.

\paragraph*{Remark (Characteristic length scale).}
The exponential cutoff introduces an intrinsic characteristic length scale (CLS)
\begin{equation}\label{eq:cls}
\ell_u=\frac{\sigma^2}{\|\drift\|}=\frac{1}{u},\quad u=\|\boldsymbol{u}\|,
\end{equation}
which suppresses long excursions parallel to the absorbing boundary and restores
finiteness of a broad class of induced information observables.
From this viewpoint, drift toward the boundary acts as a robust regularization
mechanism that interpolates between scale-free and exponentially screened exit
statistics across spatial dimensions \cite{risken1989,redner2001guide}.
The characteristic scale $\ell_u\sim u^{-1}$ delineates two asymptotic regimes.
For tangential displacements $\|\boldsymbol{r}\|\ll \ell_u$, the boundary kernel
retains the scale-free structure inherited from harmonic measure, whereas for
$\|\boldsymbol{r}\|\gg \ell_u$ the exponential factor dominates and suppresses
long excursions.
In this sense, $\ell_u$ acts as a crossover length separating diffusion-dominated
and drift-regularized behavior at the level of boundary exit laws.

\subsection{Entropy-based effective width as a geometric diagnostic}
\label{subsec:entropy_width}
As a compact diagnostic of spatial dispersion, we associate the absolutely
continuous exit law $\omega^{\xvec}$ with the differential entropy
\cite{cover2006elements}
\begin{equation}
H(\Xi\,|\,\xvec) = -\int_{\partial\Omega} \mathcal{K}(\xvec,\yvec) \log \mathcal{K}(\xvec,\yvec) \, \diff S_{\yvec}.
\end{equation}
In the zero-drift limit, the exit law reduces to the scale-free Cauchy-type
form in Eq.~\eqref{eq:general_tail_nodrift}. Its variance diverges, whereas the
differential entropy remains finite. This motivates the effective width
\begin{equation}
W_{\text{eff}} = \exp[H(\Xi\,|\,\xvec)].
\label{eq:weff_def}
\end{equation}
Physically, $W_{\text{eff}}$ measures the effective spatial footprint of the
hitting pattern. Figure~\ref{fig:FHL_localization}(a) illustrates the associated
localization of the one-dimensional exit density as drift increases. For the
same geometry with $\lambda=1$, numerical integration of the exact kernel
$\mathcal{K}^{(1)}(r;u,\lambda)$ gives the monotonic decrease of
$W_{\text{eff}}$ shown in Fig.~\ref{fig:FHL_localization}(b). Thus, the entropy-based
width provides a finite summary of the crossover from the scale-free Cauchy
regime to a drift-localized footprint governed by $\ell_u$.

This diagnostic characterizes the spread generated by a fixed point source; a
full input--output quantity such as Shannon capacity \cite{cover2006elements}
would additionally require specifying an input ensemble and lies beyond the
present scope.

\section{Curved-boundary benchmark: exterior hitting of a circle}
\label{sec:circle}
\begingroup
This section uses an exactly solvable circular receiver to organize four
complementary consequences of the planar theory within a single curved-boundary
benchmark. We first state the exact exterior-circle hitting law, which retains
global excursions around the receiver. We then zoom in near the closest boundary
point and show explicitly how the point-to-line Cauchy kernel emerges as the
local scaling limit. Next, an intermediate line is introduced to factor the
circle law into an exact Cauchy--Poisson composition. Finally, the tangent
supporting line yields a pathwise ordering and a finite-time upper bound on the
absorbed fraction. Figure~\ref{fig:circle_benchmark} summarizes these two
complementary viewpoints: panel~(a) displays the global stochastic composition,
whereas panel~(b) displays the local tangent geometry and the supporting-line
comparison. In this way, local flatness, global curvature, and pathwise geometric
constraints are treated as connected aspects of the same benchmark rather than
as separate observations.

\begin{figure*}[t]
\centering
{\color{black}
\resizebox{0.98\textwidth}{!}{%
\definecolor{receiverfill}{HTML}{DCE8F2}
\definecolor{receiveredge}{HTML}{334E68}
\definecolor{entranceblue}{HTML}{2675A6}
\definecolor{downstreamteal}{HTML}{16817A}
\definecolor{supportorange}{HTML}{C56A1A}
\definecolor{softgray}{HTML}{6B7280}
\definecolor{axisgray}{HTML}{9CA3AF}

\begin{tikzpicture}[
  >=Latex,
  line cap=round,
  line join=round,
  every node/.style={font=\small},
  paneltitle/.style={font=\small\bfseries,anchor=west},
  point/.style={circle,fill=receiveredge,inner sep=1.8pt},
  pathseg/.style={line width=1.05pt}
]

\begin{scope}[xshift=0cm]
  \node[paneltitle] at (-1.75,2.55)
    {(a) Cauchy--Poisson composition};

  \draw[->,axisgray,thin] (-1.75,0) -- (5.25,0)
    node[below left=-1pt] {$x_1$};

  \fill[receiverfill] (0,0) circle (1.35);
  \draw[receiveredge,line width=1.15pt] (0,0) circle (1.35);
  \node[receiveredge,font=\small\bfseries] at (-0.25,-0.15)
    {receiver $\mathcal C_a$};
  \fill[receiveredge] (0,0) circle (1.2pt);
  \node[below left=-1pt] at (0,0) {$O$};

  \draw[supportorange,densely dashed,line width=0.95pt]
    (1.35,-2.05) -- (1.35,2.05);
  \node[supportorange,anchor=south east] at (1.30,2.04)
    {supporting line $\mathcal P_a$};

  \draw[entranceblue,dashed,line width=1.05pt]
    (2.80,-2.05) -- (2.80,2.05);
  \node[entranceblue,anchor=south west] at (2.88,2.04)
    {intermediate line $\mathcal P_b$};

  \coordinate (X0) at (4.70,0);
  \fill[receiveredge] (X0) circle (2.2pt);
  \node[above=3pt] at (X0) {$\bm{x}_0=(R,0)$};

  \coordinate (Z) at (2.80,0.72);
  \draw[pathseg,entranceblue]
    plot[smooth] coordinates {
      (4.70,0.00) (4.38,0.28) (4.06,-0.06) (3.72,0.43)
      (3.36,0.18) (3.08,0.55) (2.80,0.72)
    };

  \coordinate (Y) at (1.064,0.831);
  \draw[pathseg,downstreamteal]
    plot[smooth] coordinates {
      (2.80,0.72) (2.54,1.11) (2.27,0.83) (2.03,1.31)
      (1.72,1.05) (1.43,1.18) (1.064,0.831)
    };

  \node[point,fill=entranceblue] at (Z) {};
  \node[entranceblue,anchor=south] at ($(Z)+(0,0.12)$)
    {$Z=(b,z)$};
  \node[point,fill=downstreamteal] at (Y) {};
  \node[downstreamteal,anchor=south east] at ($(Y)+(-0.04,0.10)$)
    {$Y=a(\cos\theta,\sin\theta)$};

  \draw[receiveredge,thin] (0,0) -- (Y);
  \draw[receiveredge,->,thin] (0.48,0) arc[start angle=0,end angle=38,radius=0.48];
  \node[receiveredge] at (0.57,0.19) {$\theta$};
  \draw[<->,receiveredge,thin] (0,-0.35) -- (1.35,-0.35)
    node[midway,below] {$a$};

  \node[entranceblue,align=center,font=\scriptsize] at (4.05,1.42)
    {entrance density\\[-1pt]$h_{R-b}(z)$};
  \node[downstreamteal,align=center] at (2.05,-1.68)
    {downstream curved-boundary law\\[-1pt]$p_{a;(b,z)}(\theta)$};

  \node[align=center,font=\small] at (1.75,-2.45)
    {$\displaystyle
      p_{a,R}(\theta)=\int_{-\infty}^{\infty}
      h_{R-b}(z)\,p_{a;(b,z)}(\theta)\,\mathrm dz$};
\end{scope}

\draw[axisgray!55,line width=0.55pt] (5.65,-2.65) -- (5.65,2.62);

\begin{scope}[xshift=8.40cm]
  \node[paneltitle] at (-1.65,2.55)
    {(b) Local limit and supporting-line bound};

  \fill[receiverfill]
    (-1.70,-2.00) --
    plot[domain=-2.00:2.00,samples=80] ({-0.18*\x*\x},{\x}) --
    (-1.70,2.00) -- cycle;
  \draw[receiveredge,line width=1.15pt]
    plot[domain=-2.00:2.00,samples=80] ({-0.18*\x*\x},{\x});
  \node[receiveredge,align=center] at (-1.17,-0.72)
    {curved\\receiver};

  \draw[supportorange,densely dashed,line width=1.05pt]
    (0,-2.10) -- (0,2.10);
  \node[supportorange,anchor=south west] at (0.06,1.73)
    {$\mathcal P_a$ (local tangent)};

  \draw[->,softgray,thin] (0,0) -- (0,1.48)
    node[left] {$s$};
  \draw[->,softgray,thin] (0,0) -- (1.02,0)
    node[below] {normal};
  \fill[receiveredge] (0,0) circle (1.5pt);
  \node[below left=-1pt] at (0,0) {nearest point};

  \coordinate (Xs) at (2.45,0);
  \fill[receiveredge] (Xs) circle (2.2pt);
  \node[above=3pt] at (Xs) {$\bm{x}_0$};
  \draw[<->,receiveredge,thin] (0,-0.48) -- (2.45,-0.48)
    node[midway,below] {$\delta=R-a$};

  \draw[downstreamteal,line width=2.2pt]
    plot[domain=0:1.08,samples=45] ({-0.18*\x*\x},{\x});
  \coordinate (Ys) at (-0.210,1.08);
  \node[point,fill=downstreamteal] at (Ys) {};
  \node[downstreamteal,anchor=east,align=right] at (-0.36,1.18)
    {local arc\\length $s$};

  \draw[pathseg,softgray]
    plot[smooth] coordinates {
      (2.45,0.00) (2.10,0.32) (1.80,0.08) (1.45,0.55)
      (1.08,0.30) (0.72,0.76) (0.32,0.55) (0,0.66)
      (-0.10,0.84) (-0.210,1.08)
    };
  \fill[supportorange] (0,0.66) circle (1.8pt);
  \node[supportorange,anchor=south west] at (0.05,0.69)
    {first hit of $\mathcal P_a$};

  \node[align=center,font=\small] at (1.20,-1.52)
    {$\displaystyle
      k_{a,a+\delta}(s)\sim
      \frac{\delta}{\pi(s^2+\delta^2)},
      \qquad \frac{\delta}{a}\ll1$};
  \node[align=center,font=\small] at (1.15,-2.25)
    {$\displaystyle
      \tau_{\mathcal P_a}\leq\tau_a
      \quad\Longrightarrow\quad
      \mathbb P(\tau_a\leq t)
      \leq
      \operatorname{erfc}\!\left(
        \frac{\delta}{\sqrt{4Dt}}
      \right)$};
\end{scope}

\end{tikzpicture}%
}}
\caption{Geometry of the curved-boundary benchmark. (a) A Brownian trajectory
starting from $\bm{x}_0=(R,0)$ first crosses the intermediate line
$\mathcal{P}_b$ at $Z=(b,z)$ before reaching the circular receiver
$\mathcal{C}_a$, illustrating the exact Cauchy--Poisson composition.
(b) Near the closest boundary point, the circular receiver is locally
approximated by its supporting tangent $\mathcal{P}_a$. The point-to-line Cauchy
kernel emerges when $\delta/a\ll1$, while the pathwise ordering
$\tau_{\mathcal{P}_a}\leq\tau_a$ yields the finite-time supporting-line bound.}
\label{fig:circle_benchmark}
\end{figure*}

Circular traps and absorbing obstacles provide canonical finite receivers in
diffusion-controlled capture problems \cite{redner2001guide,grebenkov2007spectral}.
Although the exterior-disk Poisson kernel itself is classical, we use it here
to make explicit how the planar FHL law enters a curved-receiver problem through
local scaling, stochastic composition, and a supporting-line comparison.
Let
\begin{equation}
\mathcal{C}_a
:=
\bigl\{\yvec\in\mathbb{R}^2:\|\yvec\|=a\bigr\},
\qquad a>0,
\label{eq:circle_boundary}
\end{equation}
be a circular absorbing receiver, and let the Brownian particle start from
$\xvec_0=(R,0)$ with $R>a$. We write
\begin{equation}
\tau_a
:=
\inf\bigl\{t>0:\|\mathbf{X}_t\|=a\bigr\},
\qquad
\Theta
:=
\arg\bigl(\mathbf{X}_{\tau_a}\bigr).
\label{eq:circle_hitting_variables}
\end{equation}
Planar Brownian motion is recurrent, so $\tau_a<\infty$ almost surely. The
distribution of $\Theta$ is the harmonic measure of the exterior disk and is
therefore independent of the diffusion coefficient $D$, which only rescales the
time variable \cite{doob2001classical,morters2010brownian}.

\paragraph*{Exact circular exit law.}
The conformal inversion $z\mapsto a^2/z$ maps the exterior of $\mathcal{C}_a$
to its interior and sends the starting point $R$ to $a^2/R$. Applying the
interior-disk Poisson kernel after this inversion gives
\begin{equation}
\mathbb{P}_{(R,0)}(\Theta\in\diff\theta)
=
p_{a,R}(\theta)\,\diff\theta,
\label{eq:circle_exit_measure}
\end{equation}
where
\begin{equation}
p_{a,R}(\theta)
=
\frac{R^2-a^2}
{2\pi\bigl(R^2-2aR\cos\theta+a^2\bigr)}
, 
\qquad -\pi<\theta\leq\pi.
\label{eq:circle_poisson_kernel}
\end{equation}
The kernel is normalized on $(-\pi,\pi]$ and remains strictly positive on the
back side of the receiver. Thus, the exact curved-boundary law contains global
excursions around the obstacle that cannot be reconstructed by simply retaining
the locally visible tangent patches.
For example, when $a=1$ and $R=2$, Eq.~\eqref{eq:circle_poisson_kernel} gives
$p_{1,2}(0)=3/(2\pi)$ at the nearest point and
$p_{1,2}(\pi)=1/(6\pi)$ at the opposite point, explicitly retaining a nonzero
back-side contribution.

\paragraph*{Local Cauchy limit.}
Let $\delta:=R-a$ denote the shortest source-to-boundary distance, and introduce
the local arc-length coordinate $s=a\theta$ centered at the nearest boundary
point. The corresponding density with respect to $\diff s$ is
\begin{equation}
k_{a,R}(s)
=
\frac{1}{a}\,
p_{a,R}\!\left(\frac{s}{a}\right).
\label{eq:circle_arclength_kernel}
\end{equation}
We now fix $u\in\mathbb{R}$, set $R=a+\delta$ and $s=\delta u$, and introduce
the dimensionless ratio
\begin{equation}
\varepsilon:=\frac{\delta}{a}.
\label{eq:circle_epsilon_def}
\end{equation}
Then $R=a(1+\varepsilon)$ and $s/a=\varepsilon u$. Combining
Eqs.~\eqref{eq:circle_poisson_kernel} and~\eqref{eq:circle_arclength_kernel}
gives the exact rescaled density
\begin{align}
\delta\,k_{a,a+\delta}(\delta u)
&=
\frac{\varepsilon\bigl[(1+\varepsilon)^2-1\bigr]}
{2\pi\bigl[(1+\varepsilon)^2
-2(1+\varepsilon)\cos(\varepsilon u)+1\bigr]}
\notag\\
&=
\frac{\varepsilon^2(2+\varepsilon)}
{2\pi\bigl\{\varepsilon^2
+2(1+\varepsilon)[1-\cos(\varepsilon u)]\bigr\}}.
\label{eq:circle_scaled_density_exact}
\end{align}
For fixed $u$, the cosine term satisfies
\begin{equation}
1-\cos(\varepsilon u)
=
\frac{\varepsilon^2u^2}{2}
+O(\varepsilon^4),
\qquad \varepsilon\to0.
\label{eq:circle_cos_expansion}
\end{equation}
Consequently, the denominator in
Eq.~\eqref{eq:circle_scaled_density_exact} becomes
\begin{align}
\varepsilon^2
+2(1+\varepsilon)[1-\cos(\varepsilon u)]
&=
\varepsilon^2
\bigl[1+(1+\varepsilon)u^2+O(\varepsilon^2)\bigr].
\label{eq:circle_denominator_expansion}
\end{align}
After canceling the common factor $\varepsilon^2$ between the numerator and
denominator, we obtain
\begin{equation}
\delta\,k_{a,a+\delta}(\delta u)
=
\frac{2+\varepsilon}
{2\pi\bigl[1+(1+\varepsilon)u^2+O(\varepsilon^2)\bigr]}.
\label{eq:circle_scaled_density_asymptotic}
\end{equation}
Taking the small-curvature-scale limit $\varepsilon=\delta/a\to0$ now yields
\begin{equation}
\lim_{\delta/a\to0}
\delta\,k_{a,a+\delta}(\delta u)
=
\frac{1}{\pi(1+u^2)}
.
\label{eq:circle_local_cauchy_limit}
\end{equation}
Equivalently, on the physical scale $s=O(\delta)$,
\begin{equation}
k_{a,a+\delta}(s)
=
\frac{\delta}{\pi(s^2+\delta^2)}
\bigl[1+o(1)\bigr].
\label{eq:circle_local_cauchy_physical}
\end{equation}
The leading term is precisely the point-to-line Cauchy kernel derived in
Eq.~\eqref{eq:app_2d_cauchy}. Hence the planar exit law is not merely a convenient
flat example: it is the local scaling limit of a smooth curved receiver whenever
the source-to-boundary distance is small relative to the radius of curvature.
Curvature and the global shape first enter beyond this local leading order.

\paragraph*{Exact Cauchy--Poisson composition.}
The planar kernel also acts as an exact entrance law for the curved receiver.
Choose an intermediate line
\begin{equation}
\mathcal{P}_b
:=
\bigl\{(b,z):z\in\mathbb{R}\bigr\},
\qquad a<b<R,
\label{eq:intermediate_line}
\end{equation}
which separates the source from the circle. If $Z$ denotes the vertical
coordinate of the first intersection with $\mathcal{P}_b$, then
\begin{equation}
h_{R-b}(z)
:=
\frac{R-b}{\pi\bigl[(R-b)^2+z^2\bigr]}
\label{eq:intermediate_cauchy_kernel}
\end{equation}
is the density of $Z$. Starting instead from the point $(b,z)$, the exterior
circle kernel is
\begin{equation}
p_{a;(b,z)}(\theta)
=
\frac{b^2+z^2-a^2}
{2\pi\bigl[(b-a\cos\theta)^2+(z-a\sin\theta)^2\bigr]}.
\label{eq:circle_kernel_off_axis}
\end{equation}
Every continuous path that reaches $\mathcal{C}_a$ has a unique first crossing
of $\mathcal{P}_b$. Conditioning on this crossing point and applying the strong
Markov property therefore gives the exact identity
\begin{equation}
p_{a,R}(\theta)
=
\int_{-\infty}^{\infty}
h_{R-b}(z)\,
p_{a;(b,z)}(\theta)\,\diff z
.
\label{eq:cauchy_poisson_composition}
\end{equation}
This relation may also be recognized as the half-plane Poisson representation
of the function $\xvec\mapsto p_{a;\xvec}(\theta)$, which is harmonic in the
source coordinate away from the receiver. Equation~\eqref{eq:cauchy_poisson_composition}
shows that the point-to-line law derived in the planar problem can be used as an
exact intermediate transport operator, while all downstream curvature dependence
is isolated in the second kernel.

\paragraph*{Corollary (supporting-line bound).}
The same construction gives a finite-time bound with a direct physical
interpretation. Let $\mathcal{P}_a=\{(a,z):z\in\mathbb{R}\}$ be the line tangent
to the circle at its nearest point to the source, and denote its first-hitting
time by $\tau_{\mathcal{P}_a}$. Continuity of Brownian paths implies the pathwise
ordering
\begin{equation}
\tau_{\mathcal{P}_a}\leq\tau_a.
\label{eq:supporting_line_order}
\end{equation}
Consequently, the cumulative absorbed fraction satisfies
\begin{equation}
\mathbb{P}_{(R,0)}(\tau_a\leq t)
\leq
\operatorname{erfc}\!\left(
\frac{R-a}{\sqrt{4Dt}}
\right)
, 
\qquad t>0.
\label{eq:supporting_line_cdf_bound}
\end{equation}
An equivalent Laplace-transform comparison follows by setting
$q=\sqrt{s/D}$ and solving the radial modified Helmholtz problem for the circle:
\begin{equation}
\frac{K_0(Rq)}{K_0(aq)}
\leq
e^{-(R-a)q}
, 
\qquad q>0,
\label{eq:supporting_line_laplace_bound}
\end{equation}
where the left- and right-hand sides are respectively
$\mathbb{E}_{(R,0)}[e^{-s\tau_a}]$ and
$\mathbb{E}_{(R,0)}[e^{-s\tau_{\mathcal{P}_a}}]$ \cite{Abramowitz1972}.
Although this supporting-line estimate does not give a pointwise bound on the
first-passage-time density, it provides a rigorous upper bound on the cumulative
reception probability at every finite time.

Taken together, the circle benchmark assigns three precise roles to the planar
solution: it is i) the local planar limit in this smooth curved benchmark, ii) an exact
entrance kernel in a strong Markov composition, and iii) a computable supporting-line
bound for a convex receiver. These conclusions do not amount to a closed-form
solution for arbitrary curved boundaries, but they identify concrete mechanisms
by which the planar theory remains informative beyond the half-space geometry.
\endgroup

\section{Conclusion and discussion}
\label{sec:conc}
This work has treated the first-hitting location as a boundary observable in its
own right. The generator--Green-function formulation and the conventional
boundary-flux formulation are equivalent descriptions of the same exit law:
the former identifies the induced boundary measure, and the latter gives its
density through a normal derivative. For the present Dirichlet
benchmark, this agreement validates the correspondence between an ensemble-level
description and the underlying trajectory dynamics. Placing them in a common
framework then makes their distinct analytical and pathwise structures visible
at once, without assigning exclusive status to either route.

For planar absorbing boundaries, this framework yields exact constant-drift
kernels in arbitrary ambient dimension $d$. Their asymptotics expose a sharp
organization of the exit statistics. In the zero-drift limit, the boundary law
is Cauchy-type and scale free, with algebraic tails generated by long tangential
excursions. Drift introduces exponential screening and the characteristic
length $\ell_u=\sigma^2/\|\drift\|=1/u$, thereby converting the diffuse boundary
footprint into a localized one. The two- and three-dimensional kernels agree
with Monte Carlo simulations of the underlying Langevin dynamics. The
entropy-based effective width summarizes this crossover while remaining finite
when variance ceases to be useful.

The exterior-circle calculation then clarifies which roles of
the planar solution survive in a curved geometry. The exact circular Poisson
kernel reduces locally to the point-to-line Cauchy law when the source distance
is small compared with the radius of curvature. More globally, the same planar
kernel acts as an exact entrance law in a Cauchy--Poisson composition through an
intermediate line. A tangent supporting line also gives a rigorous finite-time
upper bound on absorption by the convex receiver. Thus, local flatness, global
curvature, and pathwise ordering enter as distinct but connected aspects of one
solvable benchmark.

Taken together, these results specify both the reach and the limits of the
planar theory. A planar kernel is not a closed-form surrogate for an arbitrary
curved receiver, but it can be a local asymptotic law, an exact intermediate
transport operator, or a computable geometric bound. This distinction provides
a concrete starting point for extensions to drifted curved receivers and more
general domains. Natural next steps include combining the present
structure with boundary-integral or numerical modified-Helmholtz methods
\cite{beylkin2024representations}, extending the receiver model to partially
reactive, reflecting, or state-dependent boundary dynamics, and examining how
curvature, nonuniform or time-dependent drift, and additional transport scales
modify the induced exit measure.

\section*{Acknowledgments}
This work was supported by the National Science and Technology Council of Taiwan
under Grant No. NSTC 113-2115-M-008-013-MY3.



\appendix
\renewcommand{\thesubsection}{\arabic{subsection}}
\numberwithin{equation}{section}
\renewcommand{\theequation}{\thesection\arabic{equation}}
\titleformat{\section}{\large\bfseries\raggedright}{Appendix \thesection:}{0.75em}{}

\section{Zero-Drift Limit and Emergence of Cauchy Exit Laws}
\label{app:cauchy}
Throughout the main text, the exit law on $\partial\Omega$ is written as a boundary measure
$\omega^{\xvec}(\diff\yvec)=\mk(\xvec,\yvec)\,\diff S_{\yvec}$.
In the appendices we sometimes parameterize $\partial\Omega$ by local coordinates
$y\in\mathbb{R}^{d-1}$ (e.g., $y=r$ for $d=2$ and $y=\boldsymbol{r}$ for $d=3$),
in which case $\diff S_{\yvec}=\diff y$ and the same law is written in PDF form
$f_{\Xi}(y)=\mk(\xvec,\yvec(y))$.
More generally, for a coordinate map $y\mapsto \yvec(y)$,
the density takes the form
$f_{\Xi}(y)=\mk(\xvec,\yvec(y))\,J(y)$,
where $J(y)$ denotes the surface Jacobian.

This appendix provides a technical derivation of the zero-drift limit ($v \to 0$) for the drift--diffusion FHL laws, establishing how the boundary exit distribution converges to a multidimensional Cauchy-type law characterized by algebraic, heavy-tailed decay \cite{pandey2019}. This limiting behavior constitutes the mathematical foundation for the scale-free, diffusion-dominated regime analyzed in Secs.~\ref{sec:case} and~\ref{sec:information}. We explicitly treat planar absorbing geometries in two and three spatial dimensions---parameterized by boundary coordinates $r \in \mathbb{R}$ and $\boldsymbol{r} \in \mathbb{R}^2$, respectively, with an initial normal distance $\lambda > 0$. Notice that the planar absorbing boundaries serve as canonical local models for general smooth boundaries and illustrate the fundamental mechanism by which heavy-tailed statistics emerge from first-passage processes.

\subsection{Two-dimensional geometry: absorbing line}
In two spatial dimensions, the absorbing boundary is a line.
For a drift--diffusion process with constant drift
$\drift=(v_1,v_2)$, the boundary kernel associated with the exit law can be written
in the form
\begin{equation}
\label{eq:app_2d_kernel}
\begin{aligned}
\mk^{(1)}(r;\drift)
={}&
\frac{\|\drift\|\lambda}{\sigma^2\pi}
\exp\!\left(-\frac{v_2 \lambda}{\sigma^2}\right)
\exp\!\left(-\frac{v_1 r}{\sigma^2}\right)
\\[2pt]
&\times
\frac{
K_1\!\left(
\frac{\|\drift\|}{\sigma^2}
\sqrt{r^2+\lambda^2}
\right)
}{
\sqrt{r^2+\lambda^2}
},
\end{aligned}
\end{equation}
where $K_1(\cdot)$ denotes the modified Bessel function of the second kind.
This kernel satisfies
$\omega^{\xvec}(\diff r)=\mk^{(1)}(r;\drift)\,\diff r$
for $\xvec=(0,\lambda)$.

To examine the zero-drift regime, we take the limit $\|\drift\|\to 0$.
Using the standard asymptotic relation \cite{Abramowitz1972}:
\begin{equation}
\label{eq:app_bessel_limit}
\lim_{x\to 0} x K_1(x) = 1,
\end{equation}
the exponential prefactors in~\eqref{eq:app_2d_kernel} converge to unity, and the
leading contribution arises solely from the Bessel kernel.
A direct calculation yields
\begin{equation}
\label{eq:app_2d_cauchy}
\lim_{\|\drift\|\to 0} \mk^{(1)}(r;\drift)
=
\frac{\lambda}{\pi}\,
\frac{1}{r^2+\lambda^2}.
\end{equation}
Equation~\eqref{eq:app_2d_cauchy} coincides with the Cauchy kernel \cite{Verdu_Entropy23} on the real line
with scale parameter $\lambda$.
In particular, the exit law exhibits algebraic decay
$\mk^{(1)}(r)\sim |r|^{-2}$ and possesses no finite second moment.

\subsection{Three-dimensional geometry: absorbing plane}
In three spatial dimensions, the absorbing boundary is a plane and the exit law
is defined on $\mathbb{R}^2$.
For constant drift, the boundary kernel derived in
Sec.~\ref{sec:case} admits the representation
\begin{equation}
\label{eq:app_3d_kernel}
\begin{aligned}
\mk^{(2)}(\boldsymbol{r};\drift)
\propto{}&
\exp\!\left(
-\frac{\|\drift\|}{\sigma^2}
\sqrt{\|\boldsymbol{r}\|^2+\lambda^2}
\right)
\\[2pt]
&\times
\frac{
1+\dfrac{\|\drift\|}{\sigma^2}
\sqrt{\|\boldsymbol{r}\|^2+\lambda^2}
}{
\left(\|\boldsymbol{r}\|^2+\lambda^2\right)^{3/2}
}.
\end{aligned}
\end{equation}
In the zero-drift limit, the exponential factor converges to unity and the
remaining algebraic kernel dominates.
Consequently,
\begin{equation}
\label{eq:app_3d_cauchy}
\lim_{\|\drift\|\to 0}
\mk^{(2)}(\boldsymbol{r};\drift)
=
\frac{\lambda}{2\pi}
\frac{1}{
\left(\|\boldsymbol{r}\|^2+\lambda^2\right)^{3/2}
}.
\end{equation}
The resulting limit form corresponds to an isotropic two-dimensional Cauchy-type \cite{Verdu_Entropy23} exit law on the
boundary plane.
As in the two-dimensional case, the kernel exhibits algebraic decay
$\mk^{(2)}(\boldsymbol{r})\sim \|\boldsymbol{r}\|^{-3}$ and lacks a finite variance.

\section{Planar Joint Time--Location Law as a Consistency Check (Drift-Free Case)}
\label{app:plane_sanity}
\paragraph*{Joint vs.\ marginal.}
When a joint density $f_{\tau_\Omega,\Xi_\parallel}(t,y_\parallel)$ exists, the boundary kernel
is recovered by marginalization:
\begin{equation*}
\mk^{(0)}(y_\parallel)=\int_0^\infty f_{\tau_\Omega,\Xi_\parallel}(t,y_\parallel)\,\diff t,
\end{equation*}
consistent with $\omega^{\xvec}(\diff y_\parallel)=\mk^{(0)}(y_\parallel)\,\diff y_\parallel$.

This appendix provides a simple parabolic consistency check for the
generator--Green-function framework developed in the main text.
Throughout the paper, the exit time is treated as an internal variable and
eliminated at the level of the generator.
In the special case of a planar absorbing boundary, however, the drift-free
\emph{joint law} of the exit time and the tangential exit location admits a
classical closed-form representation \cite{redner2001guide,risken1989}.
Marginalizing over time recovers the drift-free (Cauchy-type) exit law on the
boundary, consistent with both the planar boundary kernels obtained via the elliptic route in Sec.~\ref{sec:case} and the asymptotic analysis in Sec.~\ref{sec:information}.

\subsection{Drift-free joint law in the half-space}
Let $d\ge 2$ denote the ambient dimension in this appendix  and consider the absorbing half-space
\begin{subequations}
\begin{align}
\Omega
&=\{x=(x_1,x_\parallel)\in\mathbb{R}\times\mathbb{R}^{d-1}: x_1>0\},\\
\partial\Omega
&=\{(0,y_\parallel): y_\parallel\in\mathbb{R}^{d-1}\}.
\end{align}
\end{subequations}
The process starts from $\boldsymbol{X}_0=(\lambda,x_\parallel)$ with $\lambda>0$.
Define the exit time and tangential exit location by
\begin{equation}
\tau_\Omega
:=
\inf\{t>0:\boldsymbol{X}_t\in\partial\Omega\},
\quad
\Xi_\parallel
:=
(\boldsymbol{X}_{\tau_\Omega})_\parallel\in\mathbb{R}^{d-1}.
\end{equation}

In the absence of drift, the normal and tangential components of the Brownian
motion decouple, so that the exit time is governed solely by the one-dimensional
normal motion, while tangential displacements remain unconstrained at the exit
event~\cite{morters2010brownian}.
As a consequence, the joint law of $(\tau_\Omega,\Xi_\parallel)$, written in a joint PDF form $f(\cdot,\cdot)$, factorizes as
\begin{equation}
\label{eq:app_planar_driftless_joint}
f^{0}_{\tau_\Omega,\Xi_\parallel}(t,y_\parallel)
=
f^{0}_{\tau_\Omega}(t)\,
f^{0}_{\Xi_\parallel\mid\tau_\Omega}(y_\parallel\mid t),
\quad
t>0,\ y_\parallel\in\mathbb{R}^{d-1}.
\end{equation}
Here the one-dimensional first-passage-time density in the normal direction is
\begin{equation}
\label{eq:app_planar_fp}
f^{0}_{\tau_\Omega}(t)
=
\frac{\lambda}{\sqrt{4\pi D\,t^{3}}}
\exp\!\left(-\frac{\lambda^{2}}{4Dt}\right),
\end{equation}
while the conditional tangential distribution is the free heat kernel in
$\mathbb{R}^{d-1}$,
\begin{equation}
\label{eq:app_planar_heat}
f^{0}_{\Xi_\parallel\mid\tau_\Omega}(y_\parallel\mid t)
=
\frac{1}{(4\pi Dt)^{\frac{d-1}{2}}}
\exp\!\left(
-\frac{\|y_\parallel-x_\parallel\|^{2}}{4Dt}
\right).
\end{equation}
Equation~\eqref{eq:app_planar_driftless_joint} is the planar counterpart of the
reflection-principle description of first passage \cite{redner2001guide,morters2010brownian}---the exit time is governed by
the normal motion, while tangential displacements remain unconstrained at the
exit event.

\subsection{Marginalization and recovery of the drift-free exit law}
Marginalizing~\eqref{eq:app_planar_driftless_joint} over the exit time yields the
drift-free exit law on $\partial\Omega$:
\begin{subequations}
\label{eq:app_marginal_group} 
\renewcommand{\theequation}{B6\alph{equation}} 
\begin{align}
\omega^{0}(\diff y_\parallel) &= \mathcal{K}^{(0)}(y_\parallel)\,\diff y_\parallel, \label{eq:app_marginal_def_a} \\
\mathcal{K}^{(0)}(y_\parallel) &= \int_{0}^{\infty} f^{0}_{\tau_\Omega,\Xi_\parallel}(t,y_\parallel)\,\diff t. \label{eq:app_marginal_def_b}
\end{align}
\end{subequations}
Substituting~\eqref{eq:app_planar_fp}--\eqref{eq:app_planar_heat} and defining
$\rho^2:=\|y_\parallel-x_\parallel\|^2+\lambda^2$, we obtain
\begin{equation}
\label{eq:app_marginal_integral}
\mathcal{K}^{(0)}(y_\parallel)
=
\frac{\lambda}{(4\pi D)^{\frac d2}}
\int_{0}^{\infty}
t^{-\left(1+\frac d2\right)}
\exp\!\left(-\frac{\rho^{2}}{4Dt}\right)\diff t .
\end{equation}
Using the standard change of variables $s=\rho^{2}/(4Dt)$ and the definition of the
Gamma function yields
\begin{equation}
\int_{0}^{\infty}
t^{-\left(1+\frac d2\right)}
\exp\!\left(-\frac{\rho^{2}}{4Dt}\right)\diff t
=
\left(\frac{4D}{\rho^{2}}\right)^{\frac d2}
\Gamma\!\left(\frac d2\right),
\end{equation}
and therefore
\begin{equation}
\label{eq:app_planar_cauchy}
\mathcal{K}^{(0)}(y_\parallel)
=
\frac{\lambda\,\Gamma\!\left(\frac d2\right)}{\pi^{\frac d2}}
\frac{1}{\left(\|y_\parallel-x_\parallel\|^{2}+\lambda^{2}\right)^{\frac d2}}.
\end{equation}

Equation~\eqref{eq:app_planar_cauchy} is precisely the drift-free Cauchy-type exit
law in the half-space \cite{redner2001guide,doob2001classical,grebenkov2007spectral}.
It is isotropic along the boundary and exhibits algebraic decay
$\mathcal{K}^{(0)}(y_\parallel)\sim \|y_\parallel\|^{-d}$ as
$\|y_\parallel\|\to\infty$, which is consistent with the general tail form
$\mathcal{K}^{(d-1)}(\boldsymbol{r})\sim \|\boldsymbol{r}\|^{-d}$ stated in the main text,
noting that here $y_\parallel\in\mathbb{R}^{d-1}$ parametrizes the boundary.
For $d=2$ it reduces to the standard Cauchy kernel on $\mathbb{R}$, while for
$d=3$ it coincides with the planar Poisson kernel on $\mathbb{R}^2$.
This recovers, at the parabolic (time-resolved) level, the same drift-free exit
law obtained in the main text and Appendix~\ref{app:cauchy} via the
generator--Green-function approach.

\paragraph*{Remark (scope of the consistency check).}
The purpose of this appendix is purely consistency.
It verifies, in the planar drift-free setting, that explicitly reintroducing the
time variable and marginalizing it out reproduces the same exit law as the
generator--Green-function route.
No claim is made that an equally simple joint time--location representation
persists beyond planar geometries or in the presence of general drift fields.

\section{Asymptotic Properties of Modified Bessel Functions}
\label{app:bessel}
This appendix summarizes the asymptotic behavior of the modified Bessel function of the second kind, $K_\nu(z)$. We specifically highlight the orders $\nu=1$ and $\nu=3/2$, which correspond to the boundary exit laws in two- and three-dimensional geometries, respectively. These properties underpin the transition between the diffusion-dominated and drift-regularized regimes discussed in the main text.

\subsection{Small-argument regime \texorpdfstring{($z\to 0^+$)}{(z -> 0+)}: algebraic divergence}
\label{app:bessel_small}
In the diffusion-dominated regime or at short transverse distances, the argument $z$ is small. For any fixed order $\nu > 0$, the function exhibits an algebraic divergence as $z \to 0^+$~\cite[Eq.~(9.6.9)]{Abramowitz1972}:
\begin{equation}
\label{eq:Knu_small_general}
K_\nu(z) \sim \frac{1}{2}\Gamma(\nu)\left(\frac{2}{z}\right)^\nu.
\end{equation}
For the specific cases analyzed in this work, the leading-order behaviors are:
\begin{subequations}
\label{eq:K_small_cases}
\begin{align}
    K_1(z) &\sim \frac{1}{z}, \label{eq:K1_small} \\
    K_{3/2}(z) &\sim \sqrt{\frac{\pi}{2}} z^{-3/2}. \label{eq:K3half_small}
\end{align}
\end{subequations}
These power-law divergences, $O(z^{-1})$ and $O(z^{-3/2})$, are the mathematical origin of the scale-free, heavy-tailed behavior characteristic of purely Brownian transport.

\subsection{Large-argument regime \texorpdfstring{($z\to \infty$)}{(z -> infinity)}: exponential screening}
\label{app:bessel_large}
For large arguments, which arise from significant drift or large transmission distances, $K_\nu(z)$ is dominated by exponential decay. For any fixed order $\nu$, the general expansion is given by~\cite[Eq.~(9.7.2)]{Abramowitz1972}:
\begin{equation}
\label{eq:Knu_large_general}
K_\nu(z) \sim \sqrt{\frac{\pi}{2z}} \,e^{-z} \left( 1 + \frac{4\nu^2 - 1}{8z} + \dots \right).
\end{equation}

Of particular relevance is the case $\nu=3/2$ (3D planar geometry), where the function admits an exact closed-form representation for all $z > 0$:
\begin{equation}
\label{eq:K3half_exact}
K_{3/2}(z) = \sqrt{\frac{\pi}{2z}} \,e^{-z} \left( 1 + \frac{1}{z} \right).
\end{equation}
From \eqref{eq:Knu_large_general} and \eqref{eq:K3half_exact}, the leading-order terms as $z \to \infty$ reduce to:
\begin{subequations}
\label{eq:K_large_cases}
\begin{align}
    K_1(z) &\sim \sqrt{\frac{\pi}{2z}} \,e^{-z}, \label{eq:K1_large} \\
    K_{3/2}(z) &\sim \sqrt{\frac{\pi}{2z}} \,e^{-z}. \label{eq:K3half_large}
\end{align}
\end{subequations}

The exponential factor $e^{-z}$ represents the drift-induced screening mechanism that regularizes the exit law. In the 3D planar case, the subleading $(1+z^{-1})$ factor in \eqref{eq:K3half_exact} gives rise to the polynomial correction $(1+\|\boldsymbol{u}\|\rho)$ appearing in the boundary kernel, cf.~Eq.~\eqref{eq:3Dresult-PRE}.

\begin{table}[t]
\caption{Compact dictionary between stochastic and transport terminology.}
\label{table:dictionary}
\centering
\footnotesize
\renewcommand{\arraystretch}{1.12}
\begin{tabular}{@{}p{0.42\columnwidth}p{0.50\columnwidth}@{}}
\toprule
\raggedright\textbf{Stochastic term}
& \raggedright\textbf{Transport counterpart} \tabularnewline
\midrule
\raggedright Generator $\mathcal{L}$
& \raggedright Steady advection--diffusion operator \tabularnewline
\raggedright Dirichlet Green function $\mathcal{G}(\mathbf{x},\mathbf{y})$
& \raggedright Response to a point source at $\mathbf{x}$ \tabularnewline
\raggedright Exit kernel $\mathcal{K}(\mathbf{x},\mathbf{y})$
& \raggedright Normal boundary-flux density \tabularnewline
\raggedright First-hitting location $\mathbf{X}_{\tau_\Omega}$
& \raggedright Absorption coordinate on $\partial\Omega$ \tabularnewline
\bottomrule
\end{tabular}
\end{table}

\section{Derivation of the General-Dimension Exit Kernel via the Heat-Kernel Resolvent}
\label{app:proof_dDhk}
This appendix provides a rigorous and self-contained derivation of the $d$-dimensional explicit boundary kernel presented in Eq.~\eqref{eq:dD-FHL}. Instead of postulating the free-space Green's function for the modified Helmholtz operator, we derive the boundary kernel by evaluating the resolvent of the purely diffusive heat kernel, an approach that is standard in stochastic transport.

Let $d$ denote the ambient spatial dimension. We consider the fundamental solution to the diffusion equation corresponding to the generator $\frac{\sigma^2}{2}\Delta$ in $\mathbb{R}^{d}$, which represents the transition density of a Brownian motion with diffusion coefficient $D=\sigma^2/2$ \cite{morters2010brownian,evans2010partial}. Namely, 
\begin{equation}
p(t, \xvec, \yvec)
=
\frac{1}{(2\pi\sigma^2 t)^{\frac{d}{2}}}
\exp\!\left(
-\frac{\|\xvec-\yvec\|^2}{2\sigma^2 t}
\right).
\end{equation}
To satisfy the absorbing (Dirichlet) boundary condition on the boundary $\bdry$, we use the method of images \cite{jackson1998classical}. By placing an image sink at $\xvec^*$, symmetrically opposite to the initial position $\xvec$ with respect to the boundary, the transition density of the absorbed process is
\begin{equation}
p_{\text{dir}}(t,\xvec,\yvec)
=
p(t,\xvec,\yvec)
-
p(t,\xvec^*,\yvec).
\end{equation}

For a boundary point $\yvec\in\bdry$, let $\br\in\mathbb{R}^{d-1}$ denote the tangential displacement and let $\lambda>0$ denote the initial normal distance. The distance from both the source and the image to the boundary point is
\(\rho=\sqrt{\|\br\|^2+\lambda^2}\). 
Evaluating the outward normal derivative $-\partial_{n(\yvec)}$ of $p_{\text{dir}}$ at the boundary yields
\begin{equation}
\label{eq:app_heat_flux}
-\partial_{n(\yvec)} p_{\text{dir}}(t,\xvec,\yvec)
=
\frac{2\lambda}{\sigma^2 t}
\frac{1}{(2\pi\sigma^2 t)^{\frac{d}{2}}}
\exp\!\left(
-\frac{\rho^2}{2\sigma^2 t}
\right).
\end{equation}

To connect this to the elliptic boundary-value problem formulated in Eq.~\eqref{eq:BVP-Helm_u}, we construct the Green function $\tilde{\mathcal{G}}(\xvec,\yvec)$ satisfying $(\Delta - \|\bu\|^2)\tilde{\mathcal{G}} = -\delta(\xvec-\yvec)$. Using the resolvent of the diffusion generator, this Green function is given by the time integral of the absorbed heat kernel discounted by the rate $\alpha = \frac{\sigma^2}{2}\|\bu\|^2$, and scaled by $\frac{\sigma^2}{2}$ to match the operator definition, 
\begin{equation}
\tilde{\mathcal{G}}(\xvec,\yvec) = \int_0^\infty \frac{\sigma^2}{2} e^{-\frac{\sigma^2}{2}\|\bu\|^2 t} p_{\text{dir}}(t,\xvec,\yvec) \,dt.
\end{equation}
Consequently, its outward normal derivative is
\begin{equation}
\label{eq:app_resolvent_integral}
\begin{aligned}
&-\partial_{n(\yvec)}\tilde{\mathcal G}(\xvec,\yvec) \\
&\quad = \int_0^\infty \frac{\sigma^2}{2} e^{-\frac{\sigma^2}{2}\|\bu\|^2 t} \bigl[-\partial_{n(\yvec)}p_{\text{dir}}(t,\xvec,\yvec)\bigr] \,dt \\
&\quad = \frac{\lambda}{(2\pi\sigma^2)^{\frac{d}{2}}} \int_0^\infty t^{-\frac{d+2}{2}} \\
&\qquad \times \exp\!\left( -\frac{\sigma^2}{2}\|\bu\|^2 t - \frac{\rho^2}{2\sigma^2 t} \right) dt.
\end{aligned}
\end{equation}

This integral can be evaluated analytically using the standard integral representation of the modified Bessel function of the second kind \cite{Abramowitz1972},
\begin{equation}
\int_0^\infty
\tau^{-\nu-1}
\exp\!\left(
-p\tau-\frac{q}{\tau}
\right)
d\tau
=
2\left(\frac{p}{q}\right)^{\nu/2}
K_\nu\!\left(2\sqrt{pq}\right).
\end{equation}
Substituting $\nu=\frac{d}{2}$,
$p=\frac{\sigma^2}{2}\|\bu\|^2$,
and $q=\frac{\rho^2}{2\sigma^2}$ into Eq.~\eqref{eq:app_resolvent_integral}, the argument of the Bessel function evaluates to $2\sqrt{pq} = \|\bu\|\rho$. Remarkably, the prefactor evaluates to $(p/q)^{\nu/2} = (\sigma^2\|\bu\|/\rho)^{\frac{d}{2}}$, perfectly canceling the explicit $\sigma^2$ dependence in the denominator. This yields the scale-regularized boundary Poisson kernel:
\begin{equation}
\label{eq:app_poisson_bessel}
-\partial_{n(\yvec)}\tilde{\mathcal G}(\xvec,\yvec)
=
2\lambda
\frac{\|\bu\|^{\frac{d}{2}}}{(2\pi)^{\frac{d}{2}}}
\frac{
K_{\frac{d}{2}}\!\left(
\|\bu\|\rho
\right)
}{
\rho^{\frac{d}{2}}
}.
\end{equation}

Finally, to recover the physical first-hitting location boundary kernel, we apply the exponential tilting relation established in Eq.~\eqref{eq:FHL-kernel-from-Helm_final}. The displacement vector $\yvec-\xvec$ corresponds to a tangential shift $\br$ and a normal shift $-\lambda$. Multiplying Eq.~\eqref{eq:app_poisson_bessel} by the exponential factor $\exp(\bu^{\tsp}(\yvec-\xvec)) = \exp(\bu_{\parallel}^{\tsp}\br-u_d\lambda)$ yields
\begin{equation}
\begin{aligned}
\mk^{(d-1)}(\br;\bu,\lambda)
&=
2\lambda
\frac{\|\bu\|^{\frac{d}{2}}}{(2\pi)^{\frac{d}{2}}}
\exp\!\big(\bu_{\parallel}^{\tsp}\br-u_d\lambda\big) \\
&\quad\times
\frac{
K_{\frac{d}{2}}\!\left(
\|\bu\|\rho
\right)
}{
\rho^{\frac{d}{2}}
}.
\end{aligned}
\end{equation}
This coincides exactly with the closed-form expression in Eq.~\eqref{eq:dD-FHL}. The derivation rigorously demonstrates that the high-dimensional exit law naturally resolves into a unified modified Bessel kernel structure, originating fundamentally from the integration over stochastic transport times.

\bibliographystyle{unsrt}
\bibliography{main}

\end{document}